\documentclass[letterpaper,titlepage,11pt]{article}

\usepackage{amssymb,amsmath,amsfonts,jheppub}
\usepackage{mathrsfs}
\usepackage{epsfig}
\usepackage[toc]{appendix}
\numberwithin{equation}{section}
\usepackage{graphicx}
\usepackage{subfig}
\usepackage{tikz}
\usepackage{tensor}
\usepackage{todonotes}
\usetikzlibrary{decorations.pathmorphing,calc}

\usepackage{amsmath,amssymb}
\usepackage{verbatim}
\usepackage{graphicx}
\usepackage{color}

\DeclareFontFamily{OT1}{rsfs}{}
\DeclareFontShape{OT1}{rsfs}{m}{n}{ <-7> rsfs5 <7-10> rsfs7 <10->rsfs10}{}
\DeclareMathAlphabet{\mycal}{OT1}{rsfs}{m}{n}

\newcommand{\be}[1]{ \begin{equation}\label{#1}
}

\newcommand{\bea}[1]{\begin{eqnarray}\label{#1}
}
\newcommand{\eea}{
\end{eqnarray}}

\newcommand{\tr}{\textrm{tr}}

\newcommand{\eq}[2]{\begin{equation} #1 \label{#2} \end{equation}}
\newcommand{\eps}{\varepsilon}
\newcommand{\al}{\alpha}

\newcommand{\de}{\delta}

\newcommand{\la}{\lambda}
\newcommand{\si}{\sigma}

\DeclareMathOperator{\extdm}{d}
\newcommand{\extd}{\extdm \!}

\newcommand{\vp}{\varphi}

\newcommand{\dM}{\partial{\mathcal M}}

\newcommand{\ts}[1]{\textrm{\tiny #1}}
\newcommand{\ms}[1]{\textrm{\tiny $#1$}}
\newcommand\nts{\negthickspace}
\newcommand\bns{\nts \nts \nts}
\newcommand{\EE}{\mathcal{E}}
\newcommand{\MM}{\mathcal{M}}

\newcommand\xp{\mathcal{X}^{+}}
\newcommand\xnot{\mathcal{X}^{0}}
\newcommand\xm{\mathcal{X}^{-}}

\newcommand\ap{\mathcal{L}^{+}}
\newcommand\anot{\mathcal{L}^{0}}
                       \newcommand\am{\mathcal{L}^{-}}
                       \newcommand\apm{\mathcal{L}^{\pm}}

\newcommand\veps{\varepsilon}
\newcommand\vepsp{\veps^{+}}
\newcommand\vepsnot{\veps^{0}}
\newcommand\vepsm{\veps^{-}}

\newcommand{\dd}{\mathrm{d}}

\newcommand{\mX}{{\mathbf{X}}}

\newcommand{\mytitle}{Menagerie of AdS$\boldsymbol{_2}$ boundary conditions}

\title{\mytitle}

\subheader{TUW--17--05}

\newcommand{\dg}{\ast}
\newcommand{\rmc}{\dagger}
\newcommand{\js}{\ddagger}
\newcommand{\cv}{\S}
\newcommand{\dv}{\P}

\author[\dg]{Daniel Grumiller,}
\emailAdd{grumil@hep.itp.tuwien.ac.at}
\author[\rmc]{Robert McNees,}
\emailAdd{rmcnees@luc.edu}
\author[\js]{Jakob Salzer,}
\emailAdd{salzer@hep.itp.tuwien.ac.at}
\author[\cv]{Carlos Valc{\'a}rcel,}
\emailAdd{valcarcel.flores@gmail.com}
\author[\dv]{and Dmitri Vassilevich}
\emailAdd{dvassil@gmail.com}
\affiliation[\dg,\js]{Institute for Theoretical Physics, TU Wien, Wiedner Hauptstr.~8-10/136, A-1040 Vienna, Austria}
\affiliation[\rmc]{Department of Physics, Loyola University Chicago, Chicago, IL, USA}
\affiliation[\dg,\cv,\dv]{CMCC-Universidade Federal do ABC, Santo Andr\'e, S.P. Brazil}
\affiliation[\dv]{Department of Physics, Tomsk State University, Tomsk, Russia}

\abstract{
We consider different sets of AdS$_2$ boundary conditions for the Jackiw--Teitelboim model in the linear dilaton sector where the dilaton is allowed to fluctuate to leading order at the boundary of the Poincar{\'e} disk. The most general set of boundary condtions is easily motivated in the gauge theoretic formulation as a Poisson sigma model and has an $\mathfrak{sl}(2)$ current algebra as asymptotic symmetries. Consistency of the variational principle requires a novel boundary counterterm in the holographically renormalized action, namely a kinetic term for the dilaton. The on-shell action can be naturally reformulated as a Schwarzian boundary action. While there can be at most three canonical boundary charges on an equal-time slice, we consider all Fourier modes of these charges with respect to the Euclidean boundary time and study their associated algebras. Besides the (centerless) $\mathfrak{sl}(2)$ current algebra we find for stricter boundary conditions a Virasoro algebra, a warped conformal algebra and a $\mathfrak{u}(1)$ current algebra. In each of these cases we get one half of a corresponding symmetry algebra in three-dimensional Einstein gravity with negative cosmological constant and analogous boundary conditions. However, on-shell some of these algebras reduce to finite-dimensional ones, reminiscent of the on-shell breaking of conformal invariance in SYK. We conclude with a discussion of thermodynamical aspects, in particular the entropy and some Cardyology.
\begin{center}
{\flushright\hfill\includegraphics[width=2cm]{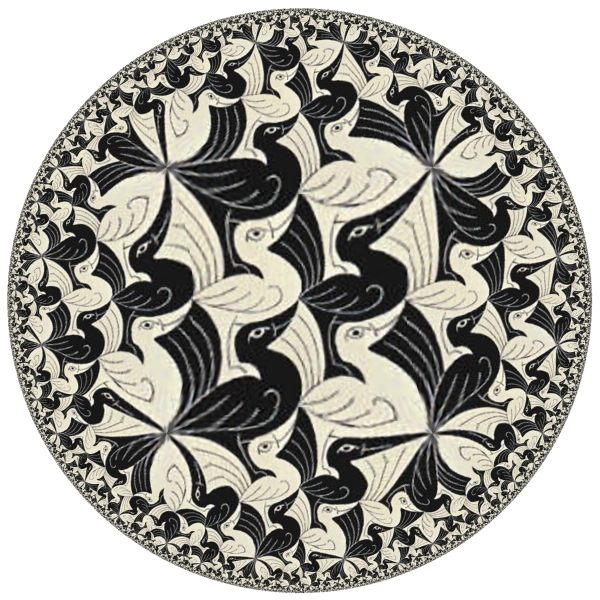}\qquad\quad\includegraphics[width=2cm]{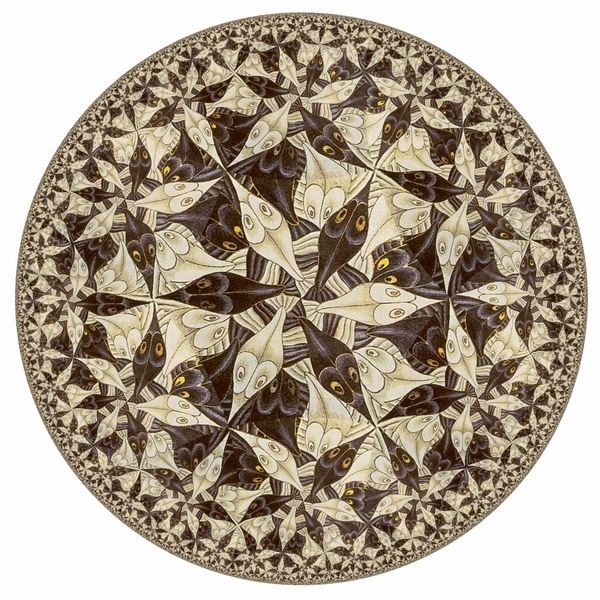}\qquad\quad\includegraphics[width=2cm]{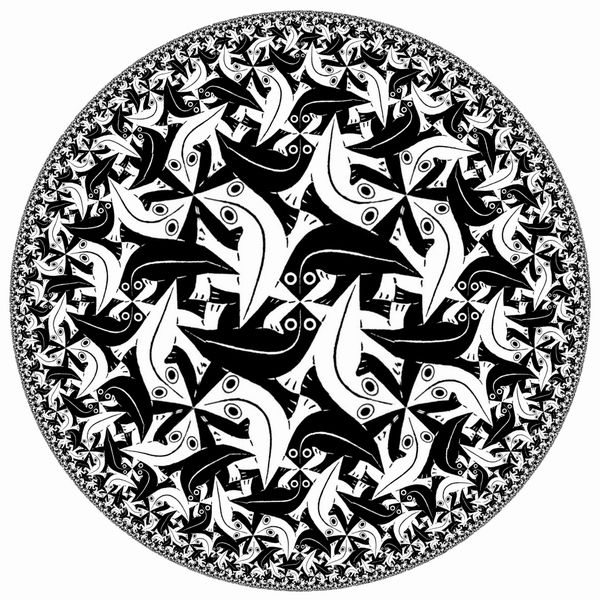}\qquad\quad\includegraphics[width=2cm]{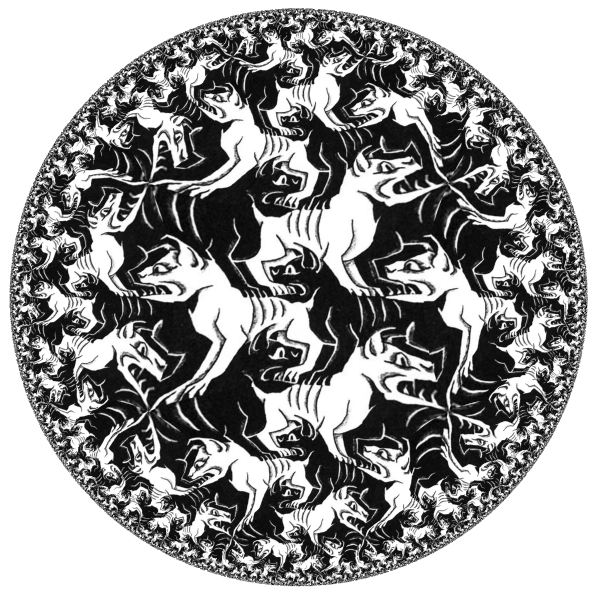}\qquad\quad\includegraphics[width=2cm]{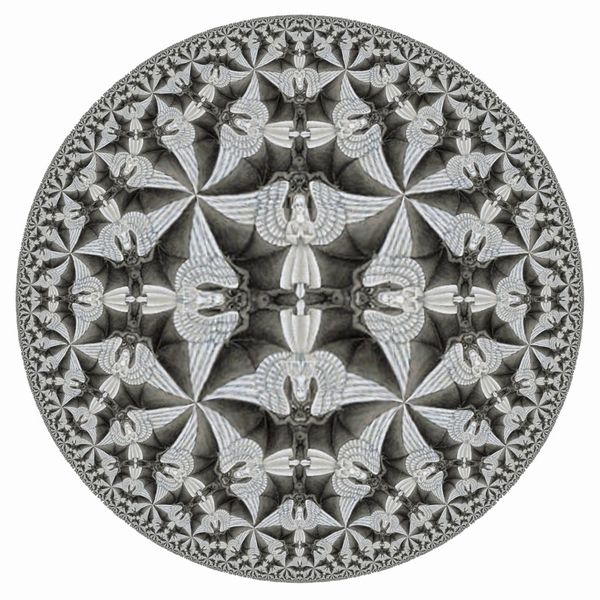}\quad\phantom{.}}
{\footnotesize Pictures created by Jos Leys at \href{http://www.josleys.com/show_gallery.php?galid=325}{http://www.josleys.com/show\_gallery.php?galid=325}, used with permission}
\end{center}
}


\begin{document}

\maketitle

\section{Introduction}\label{se:1}

The study of dilaton gravity in two dimensions began in the 1980s with the introduction of the Jackiw--Teitelboim (JT) model \cite{Jackiw:1984, Teitelboim:1984}, and has been punctuated by periods of increased interest in the community. For instance, in the early 1990s, work on the string theory black hole \cite{Mandal:1991tz, Elitzur:1991cb, Witten:1991yr, Dijkgraaf:1992ba} and the CGHS model \cite{Callan:1992rs} triggered a new round of activity and led to the emergence of a host of new models \cite{Odintsov:1991qu, Russo:1992yg}. Neglecting global effects, the path integral for all of them was calculated in \cite{Kummer:1996hy}. See the book by Brown \cite{Brown:1988} for an account of the first five years, the review \cite{Grumiller:2002nm} for a summary of the first eighteen years, and table 1 in \cite{Grumiller:2006rc} for a (non-exhaustive) list of models. 

Naturally, only after the late 1990s dilaton gravity was revisited in the context of AdS/CFT \cite{Maldacena:1997re, Gubser:1998bc, Witten:1998qj} and holographic renormalization \cite{Balasubramanian:1999re, Emparan:1999pm, deHaro:2000xn, Papadimitriou:2004ap}. Interest in AdS$_2$ holography has been re-invigorated by recent work \cite{Maldacena:2016hyu} on the Sachdev--Ye--Kitaev (SYK) model \cite{Kitaev:15ur, Sachdev:1992fk, Sachdev:2010um}.\footnote{%
For additional work related to the SYK model, see e.g.~\cite{Polchinski:2016xgd, You:2016ldz, Jevicki:2016bwu, Jensen:2016pah, Maldacena:2016upp, Engelsoy:2016xyb, Bagrets:2016cdf, Garcia-Alvarez:2016wem, Jevicki:2016ito, Gu:2016oyy, Gross:2016kjj, Berkooz:2016cvq, Garcia-Garcia:2016mno, Banerjee:2016ncu, Fu:2016vas, Witten:2016iux, Cotler:2016fpe, Klebanov:2016xxf, Davison:2016ngz, Peng:2016mxj, Krishnan:2016bvg, Turiaci:2017zwd, Ferrari:2017ryl, Bi:2017yvx, Li:2017hdt, Gurau:2017xhf, Mandal:2017thl, Gross:2017hcz, Mertens:2017mtv, Krishnan:2017txw}.
This list of references is necessarily incomplete, and we apologize for omissions.
}

The main goal of the present paper is to provide the most general AdS$_2$ boundary conditions for the JT model, to examine stricter sets of boundary conditions and their associated asymptotic symmetry algebras, and to consider their application in AdS$_2$ holography. 
Since there are numerous different approaches to AdS$_2$ holography 
we list below distinguishing features of our approach and contrast them with selected earlier work:
\begin{itemize}
    \item {\bf Linear dilaton holography.} There is a simple, though incorrect, argument that the linear dilaton sector cannot have asymptotic symmetries containing the AdS$_2$ isometries, $sl(2,\mathbb{R})$. Namely, only one of the three AdS$_2$ Killing vectors $\xi$ is capable of obeying ${\cal L}_\xi X = \xi^\mu\partial_\mu X = 0$ if the dilaton $X$ is not constant. While this observation is correct, it does not imply that the asymptotic symmetries (which are not necessarily isometries) cannot contain the AdS$_2$ isometries --- they just cannot be isometries of the combined metric and dilaton system. Instead, they must transform the dilaton in a non-trivial way. If this is allowed by the boundary conditions then the AdS$_2$ isometries remain part of the asymptotic symmetries. By contrast, the constant dilaton sector maintains all AdS$_2$ isometries as asymptotic symmetries in a more obvious way. Partly for this reason, partly for simplicity, and partly because constant dilaton vacua naturally emerge in near horizon extremal geometries \cite{Kunduri:2007vf}, most of the early literature on AdS$_2$ holography focussed on the constant dilaton sector \cite{Strominger:1998yg, Maldacena:1998uz, Brigante:2002rv, Astorino:2002bj, Verlinde:2004gt, Gupta:2008ki, Alishahiha:2008tv, Sen:2008vm, Hartman:2008dq, Castro:2008ms, Balasubramanian:2009bg, Castro:2009jf}.
    \item {\bf Black hole holography.} This point is related to the previous one. For constant dilaton solutions, while there exists a horizon, it does not make sense to interpret the corresponding spacetime as a black hole. Even though these spacetimes formally have an entropy proportional to the value of the dilaton at the horizon, $X_h$, this value is not well-defined since translation invariance allows to shift $X_h\to X_h+\rm const$.  Also the evaluation of the quantum partition function reveals that there is only one physical state for constant dilaton vacua \cite{Grumiller:2015vaa}.  By contrast, for linear dilaton solutions there is a unique center, namely the point where the effective Newton constant (given by the inverse of the dilaton) tends to infinity. The existence of such a center breaks translation invariance and gives the value of entropy an operational meaning. Also the partition function turns out to be non-trivial for linear dilaton solutions \cite{Grumiller:2007ju}. Thus, using the linear dilaton sector also allows to address aspects of holography of black holes in two dimensions.
    \item {\bf Fluctuating dilaton holography.} Earlier work on linear dilaton holography imposed Dirichlet boundary conditions on the dilaton, see \cite{Grumiller:2007ju, Cvetic:2016eiv} and references therein. By contrast, in the present work we do not impose such conditions and instead let the dilaton fluctuate to leading order at the boundary.
    \item {\bf Euclidean Poincar{\'e} holography.} To the best of our knowledge, the first holographic approach allowing for a fluctuating dilaton at the boundary was \cite{Grumiller:2013swa}. However, in that paper a well-defined variational principle was only achieved in the presence of a disconnected boundary (i.e., global AdS$_2$ with topology of a strip) so that non-vanishing variations from one boundary component are cancelled by a similar contribution, but with opposite sign, from the other boundary component. By contrast, in the present work we put the Lobachevsky plane on the Poincar{\'e} disk and thus have an $S^1$ as boundary (we work in Euclidean signature, so the ``angular'' coordinate $\vp$ of the disk is Euclidean time and its periodicity related to inverse temperature).
    \item {\bf Pure dilaton gravity holography.} Many earlier papers considered an additional Maxwell field, whose presence was often crucial to provide non-trivial features of the model (e.g.~to provide a constant dilaton solution \cite{Hartman:2008dq} or a state-dependent cosmological constant \cite{Grumiller:2014oha}), or additional (scalar) matter fields (e.g.~to provide a carrier of Hawking quanta, to address black hole evaporation, scattering and backreactions \cite{Callan:1992rs, Kummer:1998zs, Kummer:1999zy, Grumiller:2000ah, Fischer:2001vz, Almheiri:2014cka}). By contrast, in the present work we stick to pure dilaton gravity without a Maxwell field (though our discussion readily generalizes to the case with gauge fields) and without extra matter fields.
    \item {\bf General boundary conditions.} All previous approaches considered a metric that was fixed at the boundary. By contrast, in our most general setup the metric is allowed to fluctuate to leading order at the AdS$_2$ boundary. Restricting the metric in various ways then leads to different sets of boundary conditions with corresponding sets of asymptotic symmetries. In essence, our discussion of boundary conditions follows the AdS$_3$ discussion in \cite{Grumiller:2016pqb}.
    \item {\bf Canonical boundary charges.} Most previous approaches either disregarded the canonical boundary charges, concluded that they vanish (which for many cases, such as constant dilaton, is true) or found non-integrable results, see \cite{Grumiller:2015vaa} and refs.~therein. For the boundary conditions discussed in this work we find finite, non-trivial and integrable charges.
\end{itemize}

Actually, let us expand on the last point. As we shall review, standard canonical analysis leads to at most three canonical boundary charges $Q^I(\vp)$ that may depend on the boundary coordinate $\vp$. This necessarily implies that the canonical realization of the asymptotic symmetry algebra leads to finite dimensional algebras like $u(1)$; in particular, no Virasoro or current algebra can emerge in this way, which means that one cannot apply Cardyology to AdS$_2$ black holes in any obvious way. As we shall demonstrate, off-shell (and in some cases also on-shell) some of the charges have an infinite tower of Fourier modes with respect to $\vp$. 

Geometrically, this is related to the $\vp$-dependence of the dilaton field at the boundary $S^1$. Since there is always a diffeomorphism mapping a non-round $S^1$ to the round one, one could conclude that all such $\vp$-dependence is spurious. Whether or not this is the right conclusion depends on whether or not these diffeomorphisms are proper ones. Actually, this question is analogous to its higher-dimensional version of whether or not near horizon soft hair \cite{Hawking:2016msc} is associated with physical states. At least for the boundary conditions introduced in \cite{Donnay:2015abr, Afshar:2016wfy} it turns out that the diffeomorphisms that map the non-round horizon $S^1$ of black flowers to a round $S^1$ of black holes (``soft boosts'') are not proper ones as they do change the physical state.  

For similar reasons, in the present work we keep all the Fourier modes of the charges $Q^I(\vp)$. Technically, it is then useful to work with ``time-averaged charges'', i.e., charges integrated over the boundary $S^1$, since this allows to partially integrate and drop total derivative terms (this trick was already used in \cite{Cadoni:1999ja}). We then study the algebra associated with all these Fourier modes and find sets of boundary conditions where this algebra is infinite dimensional on-shell. In the remaining cases the algebra looks infinite-dimensional off-shell, but reduces to a finite-dimensional subalgebra on-shell.

\enlargethispage{0.5truecm}
We summarize now the main results that we find by implementing the approach outlined above.
\begin{itemize}\setlength{\itemsep}{-0.1ex}
    \item We present the loosest set of boundary conditions, \eqref{eq:ica}-\eqref{eq:b1}, for the JT model (in first order formulation), leading to a generalized Fefferman--Graham expansion of metric \eqref{AdSLineElement} and dilaton \eqref{AdSDilaton}.
    \item We show that the variational principle is well-defined in the first order formulation, provided a certain quantity (denoted by $1/Y$) has a fixed zero-mode \eqref{eq:FirstOrderGammaVariation}.
    \item We provide a pair of counterterms, supplementing the action of \cite{Grumiller:2007ju} with a boundary kinetic term for the dilaton, that holographically renormalize the second order action \eqref{2ndOrderAction}. Consistency of the variational principle again requires $1/Y$ to have a fixed zero mode \eqref{eq:OnShellVariation}.
    \item The asymptotic symmetries (i.e., diffeomorphisms that leave the action invariant modulo diffeomorphisms that reduce to identity at the boundary), see \eqref{DiffeoTransformhp}, coincide with the large gauge transformations in the Poisson sigma model formulation \eqref{eq:lgt}, which form an $\mathfrak{sl}(2)$ current algebra for the loosest set of boundary conditions.
    \item The on-shell action reduces in general to dAFF conformal quantum mechanics \cite{deAlfaro:1976je} at the boundary \eqref{eq:40}.
    \item Eliminating a term from the on-shell action in conflict with homogeneity in the dilaton allows to represent the on-shell action as Schwarzian action \eqref{eq:43}.
    \item The canonical realization of the asymptotic symmetries (in the sense explained above) leads to a centerless $\mathfrak{sl}(2)$ current algebra for the loosest set of boundary conditions.
    \item Fixing the leading order boundary metric and demanding time-inversion invariance leads to conformal boundary conditions, where the remaining function appearing in the metric transforms under a diffeomorphism of the boundary by an infinitesimal Schwarzian derivative. The final result for the charges also leads to a Virasoro algebra off-shell, but only to a single generator (essentially the mass) on-shell \eqref{eq:43a}. Thus, like in SYK we have an on-shell breaking of conformal symmetry.
    \item Keeping the assumption of time-inversion invariance but allowing the leading order boundary metric to fluctuate enhances the Virasoro to a warped conformal algebra \eqref{eq:warpedcoad} or \eqref{eq:nicewcft}. Otherwise, we recover the same SYK-like features discussed above.
    \item The final set of boundary conditions leads to a $\mathfrak{u}(1)$ current algebra \eqref{eq:u1} (both off-shell and on-shell). 
    \item Our on-shell action leads to the correct free energy and entropy, compatible with the first law, see \eqref{eq:39}-\eqref{eq:S}.
    \item Finally, we employ some Cardyology to recover the macroscopic result of entropy from a chiral Cardy formula \eqref{eq:cardy}. A key result here is the non-zero central charge $c=6k\bar Y/\pi$ derived in \eqref{eq:43b}, where $k$ is essentially the inverse two-dimensional Newton constant and $1/\bar Y$ the zero mode of $1/Y$. The reason our result is not in conflict with the usual lore that two-dimensional quantum gravity must have $c=0$ (see e.g.~\cite{Knizhnik:1988ak, David:1988hj, Distler:1989jt}) is that our central charge (and the associated Virasoro algebra) appears only off-shell, but not on-shell. We also match with the warped conformal entropy formula \eqref{eq:warped2}.
\end{itemize}
\enlargethispage{0.5truecm}

This paper is organized as follows. In section \ref{se:0} we introduce the JT model in the gauge theoretic formulation as Poisson sigma model. In section \ref{se:3} we present our loosest set of boundary conditions in the first order formulation. In section \ref{se:2nd} we translate our results into second order formulation and holographically renormalize the action to show that we have a well-defined variational principle. In section \ref{sec:varprinc} we make contact with SYK and derive the Schwarzian action, as well as conformal quantum mechanics. In section \ref{se:2} we discuss the canonical boundary charges for several different sets of boundary conditions. In section \ref{se:4} we address thermodynamics and entropy of AdS$_2$ black holes. We also show that naive Cardyology works, not just in the Virasoro case but also for warped conformal boundary conditions, in the sense that the chiral Cardy formula (and its warped conformal generalization) lead to a result for entropy compatible with the macroscopic Wald entropy. In section \ref{se:5} we conclude. Appendix \ref{app:ModifiedBracket} provides the relation between asymptotic symmetries in first and second order formulations. 
Appendix \ref{app:B} discusses toy models that amount to Poisson sigma models in Casimir--Darboux coordinates, which elucidates some of the subtle issues encountered in the main text and paves the way towards generic models of two-dimensional dilaton gravity.

\section{Jackiw--Teitelboim as Poisson sigma model}\label{se:0}

Before starting we mention some of our conventions. We work in Euclidean signature throughout and assume that our two-dimensional manifold has the topology of a disk. Where applicable, we use the notations and conventions of \cite{Grumiller:2015vaa, Grumiller:2016dbn}.

Like in three dimensions, where the first order formulation as Chern--Simons theory \cite{Achucarro:1986vz, Witten:1988hc} has many technical advantages, also in two dimensions the first order formulation is useful. In this section we briefly summarize this formulation, with particular focus on the JT model.

The bulk action for the JT model in the first order form reads
\begin{equation}
I_{\rm JT}=-\frac k{2\pi} \int \big( X^a (\dd e_a + \epsilon_a^{\ b}\omega\wedge e_b)+X\dd \omega + \tfrac 12 X
\epsilon^{ab} e_a\wedge e_b\big) ~. \label{JT1st}
\end{equation}
The indices $a, b$ take values $0$ and $1$, with $e_{a}$ the vielbein and $X^{a}$ a pair of Lagrange multipliers enforcing the torsion constraint. The dilaton itself is denoted by $X$, and $\omega$ is the (dualized) spin connection. Forming the triplet $X^{I} = \{X^{0}, X^{1}, X\}$ from the dilaton and Lagrange multipliers, and collecting the vielbein and spin connection one-forms together in $A_{I} = \{e_{0}, e_{1}, \omega\}$, the JT model can be written as a Poisson sigma model (PSM) \cite{Schaller:1994es,Ikeda:1994fh}
\begin{equation}
  \label{eq:1}
  I_{\rm JT}=-\frac{k}{2\pi}\int\big(X^I\dd A_I+\tfrac{1}{2}P^{IJ}(X^K)\, A_I\wedge A_J\big)\, ,
\end{equation}
with Poisson tensor ($P^{IJ}=-P^{JI}$, $P^{IL}\partial_L P^{JK} + \textrm{cycl}(I,J,K)=0$)
\begin{equation}
  \label{eq:2}
  P^{Xb}=X^{a}\epsilon\indices{_a ^b}\qquad\qquad P^{ab}=X\epsilon^{ab}\, .
\end{equation}
Extremizing the PSM action yields the equations of motion
\eq{
  \dd X^I+P^{IJ}A_J=0\qquad\qquad \dd A_I+\frac{1}{2}\partial_IP^{JK}A_J\wedge A_K=0\,.
}{eq:31}
Because the Poisson tensor is linear in the $X^{I}$, it follows from the second equation of motion that the action \eqref{eq:1} vanishes on-shell. If the Poisson tensor also vanishes on-shell (for JT this means $X=X^a=0$) we have a constant dilaton solution, otherwise a linear dilaton solution.

The PSM action is exactly invariant under the non-linear gauge transformations
\begin{equation}
  \label{eq:3}
  \delta_\lambda X^I=P^{IJ}\lambda_J\qquad \delta_\lambda A_I=-\dd \lambda_I-\partial_IP^{JK}\lambda_KA_J\, ,
\end{equation}
where $\lambda_{I}$ is a triplet of gauge parameters.
Introducing a metric $\eta_{IJ} = \text{diag}(+1,+1,-1)$ on the target space, with volume form $\epsilon^{01X} = 1$, the (linear) gauge transformations for the JT model can be expressed as
\begin{equation}
  \label{eq:4}
  \delta_\lambda X^{I}=\epsilon^{IJK}\lambda_JX_K\qquad \delta_\lambda A_I=-\dd \lambda_I-\epsilon_{IJK}A^J\lambda^K\,.
\end{equation}
Now choose $\mathfrak{so}(2,1)$ generators $J_I$ satisfying the algebra
\begin{equation}
  \label{eq:6}
  [J_0,J_1]=J_X\qquad [J_1,J_X]=-J_0\qquad [J_X,J_0]=-J_1\, ,
\end{equation}
with invariant bilinear form given by
\begin{equation}
  \label{eq:15}
  \langle J_IJ_J\rangle=\frac{1}{2}\eta_{IJ}\,.
\end{equation}
Then in terms of the Lie-algebra valued quantities $\mX=X^IJ_I$, $A=A_IJ^I$, and $\lambda=\lambda_IJ^I$, the transformations are
\begin{equation}
  \label{eq:5}
  \delta_\lambda \mX=\left[\lambda,\mX\right]\qquad \delta_\lambda A=-D\lambda\equiv-(\dd\lambda+[A,\lambda])\,.
\end{equation}
\par

It will be convenient to pass back and forth between $\mathfrak{so}(2,1)$ and $\mathfrak{sl}(2)$ bases for the fields. The transformation to $\mathfrak{sl}(2)$ generators is given by
\begin{equation}
  \label{eq:7}
  L_0=J_1\qquad L_+=J_0+J_X\qquad L_-=J_X-J_0\,,
\end{equation}
with inverse transformation
\begin{equation}
  \label{eq:8}
  J_X=\frac{1}{2}(L_++L_-)\qquad J_0=\frac{1}{2}(L_+-L_-)\,.
\end{equation}
The $\mathfrak{sl}(2)$ generators obey the commutation relations
\begin{equation}
  \label{eq:9}
  \left[L_I,L_J\right]=(I-J)L_{I+J}\,\qquad I,J=+1,-1,0 \, ,
\end{equation}
with invariant bilinear form given by
\begin{equation}
  \label{eq:21}
  \langle L_IL_J\rangle=\kappa_{IJ}\equiv\begin{pmatrix}0&0&-1\\0&1/2&0\\-1&0&0\end{pmatrix}_{IJ}\,.
\end{equation}
Thus, the action \eqref{eq:1} can also be written as
\begin{equation}
  \label{eq:22}
  I=\frac{k}{\pi}\int_{\mathcal M}\textrm{tr}(\mX (\dd A+A\wedge A)) ~,
\end{equation}
with equations of motion
\begin{subequations}
\begin{align}
  \label{eq:32}
  DA&=\dd A+A\wedge A=\dd A+\frac{1}{2}A^I\wedge A^J[L_I,L_J]=0\\
\label{eq:32b}
  D\mX&=\dd \mX+[A,\mX]=0\,.
\end{align}
\end{subequations}
Throughout most of this paper, we work in the $\mathfrak{sl}(2)$ basis. Since both bases have an element labeled ``$0$'', we henceforth use hatted indices $\hat{0}$ and $\hat{1}$ for components in the $\mathfrak{so}(2,1)$ basis. The dictionary relating the components of the fields to the geometric variables of \eqref{JT1st} is
\begin{subequations}
     \label{eq:trans}
\begin{align}
e_{\mu \hat 0}&=A_\mu^+-A_\mu^- & X^{\hat 0}&=X^+-X^-\\
e_{\mu \hat 1}&=A_\mu^0 & X^{\hat 1}&=X^0\\
\omega_\mu&=-A_\mu^+-A_\mu^- &X&=X^++X^- \,.
\end{align}
\end{subequations}
The components appearing on the right-hand-side of these formulas correspond to the coefficients of $L^{\pm,0}$ in the $\mathfrak{sl}(2)$ valued fields $\mX$ and $A$. By the hatted indices $\hat{0}$ and $\hat{1}$ we denote the components in the $\mathfrak{so}(2,1)$ basis.

\section{Action and degrees of freedom}\label{se:3}

In this section we identify the boundary degrees of freedom in the JT model, describe the mapping to AdS$_2$ asymptotics, and establish a non-trivial action for the theory in the first order formalism.

\subsection{Auxiliary asymptotic conditions}\label{se:3-1}

We denote the coordinate along the boundary by $\varphi$ and the bulk coordinate by $\rho$. We will consider spacetimes $\mathcal{M}$ with the topology of a disk (or possibly of a cylinder\footnote{For the Hamiltonian reduction of generic PSMs on the cylinder, see \cite{Vassilevich:2013ai}.}, after suitable modifications) such that the boundary $\dM$ is $S^1$.

Let us consider an auxiliary asymptotic value problem where all fields tend to some finite values on the boundary. In the $\mathfrak{sl}(2)$ basis, the boundary values of the canonical variables may be written in a $\rho$-independent way as
\begin{align}
a_\vp(\varphi) &= L_+ {\cal L}^+(\vp) + L_0 {\cal L}^0(\vp) + L_- {\cal L}^-(\vp) \label{eq:ica}\\
x(\varphi) &= L_+ {\cal X}^+(\vp) + L_0 {\cal X}^0(\vp) + L_- {\cal X}^-(\vp) \label{eq:icx}\,.
\end{align}
They have to satisfy the equations of motion that follow from varying $A_\rho$ in the action
\begin{subequations}
\label{eq:psmeom}
\begin{align}
  \label{eq:34}
  (\mathcal{X}^\pm)^\prime\pm(\mathcal{L}^\pm\mathcal{X}^0-\mathcal{L}^0\mathcal{X}^\pm)&=0\\
\label{eq:34b}
(\mathcal{X}^0)^\prime+2(\mathcal{L}^+\mathcal{X}^--\mathcal{L}^-\mathcal{X}^+)&=0\,.
\end{align}
\end{subequations}
Here (and elsewhere) the prime denotes a derivative with respect to $\varphi$. The ansatz for the boundary values \eqref{eq:ica}, \eqref{eq:icx} is the most general one possible. 

Taken together, the equations imply that a particular combination of the $\mathcal{X}^{I}$ is constant. If we define the Casimir $\mathcal{C}$ as
\begin{gather}\label{eq:Casimir}
	\mathcal{C} = \mathcal{X}^{+} \mathcal{X}^{-} - \frac{1}{4}\,(\mathcal{X}^{0})^2
\end{gather}
then $\mathcal{C}'=0$.
The remaining equations may be enforced by solving for, say, $\mathcal{L}^{0}$ and $\mathcal{L}^{-}$ in terms of $\mathcal{L}^{+}$ and the $\mathcal{X}^{I}$. It is convenient to parameterize $\mathcal{L}^{+}$ in terms of its ratio with $\mathcal{X}^{+}$, so that
\begin{eqnarray} \label{LplusXplusRatio}
&&\mathcal{L}^{+}=\frac{1}{Y}\,\mathcal{X}^{+} \label{eq:Y}\\
&&\mathcal{L}^{0}=\frac{1}{Y}\,\mathcal{X}^{0}+\frac {(\mathcal{X}^+)'}{\mathcal{X}^+} \\
&&\mathcal{L}^{-}=\frac{1}{Y}\,\mathcal{X}^{-}+\frac {(\mathcal{X}^{0})'}{2\mathcal{X}^+}
\end{eqnarray}
where $Y = \xp/\ap$ is, for the moment, an arbitrary function of $\vp$.

If the gauge parameter $\lambda$ does not vanish at $\dM$, the corresponding gauge transformations  act on the asymptotic values of the fields. In the $\mathfrak{sl}(2)$ basis the components of $\lambda$ at the boundary are
\begin{gather}\label{eq:BoundaryGaugeTransformation}
	\lambda|_{\dM} \equiv \varepsilon(\varphi) =  L_{+}\,\varepsilon^{+}(\varphi) + L_{0}\,\varepsilon^{0}(\varphi) + L_{-}\,\varepsilon^{-}(\varphi) ~.
\end{gather}
The response of the fields to such a transformation is given by
\begin{subequations}
\label{eq:lgt}
\begin{align}
	\delta_{\varepsilon} \mathcal{L}^\pm &=  \pm\varepsilon^\pm\,\mathcal{L}^{0} \mp \varepsilon^{0}\,\mathcal{L}^\pm - \varepsilon^{\pm}{}' &
	\delta_{\varepsilon} \mathcal{L}^{0} &=  2\,\varepsilon^{+}\,\mathcal{L}^{-} - 2\,\varepsilon^{-}\,\mathcal{L}^{+} - \varepsilon^{0}{}' 
	\\
	\delta_{\varepsilon} \mathcal{X}^\pm &=  \pm\varepsilon^\pm\,\mathcal{X}^{0} \mp \varepsilon^{0}\,\mathcal{X}^\pm  &
	\delta_{\varepsilon} \mathcal{X}^{0} &=  2\,\varepsilon^{+}\,\mathcal{X}^{-} - 2\,\varepsilon^{-}\,\mathcal{X}^{+}  
	\,. \label{eq:deX}
\end{align}
\end{subequations}
The Casimir $\mathcal{C}$ is invariant, while the ratio $1/Y$ appearing in \eqref{LplusXplusRatio} transforms as a total derivative plus a term proportional to the equation of motion \eqref{eq:34}:
\begin{gather}\label{eq:OtherTransformations}
	\delta_{\varepsilon}\mathcal{C} = 0 \qquad\qquad \delta_{\varepsilon}\Big(\frac{1}{Y}\Big) = -\partial_{\varphi}\Big( \frac{\varepsilon^{+}}{\mathcal{X}^{+}}\Big) - \vepsp\,\frac{(\xp)'+\ap\,\xnot - \anot\,\xp}{(\xp)^{2}} ~.
\end{gather}
The transformations of $\mathcal{X}^I$ and $\mathcal{L}^{I}$ are identical to those of an $\mathfrak{sl}(2)$ current, with an anomalous term in the case of the $\mathcal{L}^{I}$. As we shall see below, the canonical charges are constructed from $\mathcal{X}^I$.

\subsection{Asymptotic AdS\texorpdfstring{$_2$}{2} conditions}\label{se:3-2}

The auxiliary asymptotic conditions \eqref{eq:ica}, \eqref{eq:icx} may be mapped to asymptotic  AdS$_2$ conditions by means
of the transformation (c.f. \cite{Grumiller:2013swa})
\begin{align}
 A &= b(\rho)^{-1}\,\big(\extd + a_\vp(\vp)\extd \varphi\big)\,b(\rho) \label{eq:aca} \\
 \mX &= b(\rho)^{-1}\,x(\vp)\,b(\rho)\label{eq:acx}
\end{align}
with some group element $b$ that depends only on the ``radial'' coordinate $\rho$. Equations
\eqref{eq:aca} and \eqref{eq:acx} fix the asymptotic
form of the fields. The choice of the group element is irrelevant for the gauge theoretic interpretation of the theory and only becomes relevant for a geometric interpretation. Throughout this paper we fix the group element to allow an asymptotic AdS$_2$ interpretation of our results. A convenient choice in most (but not all) cases is
\eq{
b = e^{\rho L_0} \,.
}{eq:b1}
For arbitrary boundary data $\mathcal{L}^{\pm,0}$ and $\mathcal{X}^{\pm,0}$, and using the Baker--Campbell--Hausdorff identities $b^{-1}L_{\pm}b = L_{\pm} e^{\pm\rho}$, we then obtain
\begin{gather}
A=\left(\extd \rho+\mathcal{L}^{0}\extd\varphi\right)L_{0}+\left(e^{\rho}\mathcal{L}^{+}L_{+}+e^{-\rho}\mathcal{L}^{-}L_{-}\right)d\varphi \label{L104} \\
\mX = e^{\rho}\,\mathcal{X}^{+} L_{+} + \mathcal{X}^{0} L_{0} + e^{-\rho}\,\mathcal{X}^{-} L_{-} \, .
\end{gather}

From these formulas one can easily extract the zweibein and the spin connection
\begin{align}
e_{\rho \hat 0}&=0 & e_{\rho \hat 1}&=1 & \omega_{\rho}&=0\label{L106a}\\
e_{\varphi \hat 0}&=e^{\rho}\mathcal{L}^{+}-e^{-\rho}\mathcal{L}^{-} &
e_{\varphi \hat 1}&=\mathcal{L}^{0} &
\omega_{\varphi}&=-\left(e^{\rho}\mathcal{L}^{+}+e^{-\rho}\mathcal{L}^{-}\right)\label{L106b}
\end{align}
which corresponds to the asymptotically AdS$_{2}$ line element
\begin{gather}\label{AdSLineElement}
	\extd s^2 = \extd\rho^2 + 2\,\mathcal{L}^{0} \extd\rho \extd\varphi + \Big(\left(e^{\rho}\mathcal{L}^{+}-e^{-\rho}\mathcal{L}^{-}\right)^{2}+\left(\mathcal{L}^{0}\right)^{2}\Big)\,\extd\varphi^2 ~.
\end{gather}
The generalized Fefferman--Graham expansion of the metric \eqref{AdSLineElement} 
is reminiscent of its AdS$_3$ version \cite{Grumiller:2016pqb}. Likewise, the dilaton is
\begin{gather}\label{AdSDilaton}
	X = e^{\rho}\,\mathcal{X}^{+} + e^{-\rho}\,\mathcal{X}^{-} ~.
\end{gather}
Notably, the leading order coefficients in the metric, ${\cal L}^+$, and dilaton, ${\cal X}^+$, are allowed to fluctuate. These fluctuations are almost independent from each other; as we shall explain in the next subsection, the ratio ${\cal L}^+/\xp=1/Y$ must have a fixed zero mode.

\subsection{Action in first order formalism}\label{se:3-3}

The action \eqref{JT1st} reproduces the equations of motion for the JT model, but it is trivial in the sense that it is exactly zero for any solution of those equations. We wish to find an action that yields the same equations of motion and is invariant under the same gauge transformations, but is also generically non-zero when evaluated on solutions. 

The action \eqref{eq:1} is equivalent to the `bulk' term in the more familiar second-order formulation of the JT model. So let us supplement it with a number of boundary terms to give an action of the form
\begin{align}\label{eq:JT1}
	\Gamma = & \,\, I_\ts{JT} + \frac{k}{2\pi}\,b_{0}\int_{\dM} \extd\varphi \, \Big( X \omega_{\vp } + \frac{X}{A_{\vp\hat 0}^{2}+A_{\vp\hat 1}^{2}}\,(A_{\vp\hat 0}\,A_{\vp\hat 1}' - A_{\vp\hat 1}\,A_{\vp\hat 0}') \Big)
\\ \nonumber
 & \,\, + \frac{k}{2\pi}\,b_1 \int_{\dM} \extd\varphi\,\Big(X^{a}\,A_{\vp a} +  L_\ts{CT}(x,a) \Big)~,
\end{align}
where $b_0$ and $b_1$ are constants, and $L_\ts{CT}$ is some function of the boundary values of the fields. The first boundary term in \eqref{eq:JT1}, with coefficient $b_0$, is the Gibbons--Hawking--York (GHY) term expressed in terms of the $\mathfrak{so}(2,1)$ components of the bulk fields. Our initial expectation is that some values for the constants $b_0$ and $b_1$ and an appropriate choice of $L_\ts{CT}$ will give an action that admits a well-defined variational principle for the boundary conditions \eqref{eq:ica}, \eqref{eq:icx}, possibly subject to further restrictions.

The first variation of the action should vanish when evaluated on solutions of the equations of motion, for all field variations that keep some as-yet-undetermined combination of the fields fixed on $\dM$. Here we take $\dM$ to be the $\rho \to \infty$ limit of a constant $\rho$ surface. This immediately places a condition on $b_0$ and $b_1$. The variation of the bulk term \eqref{eq:1} gives a boundary integral of $X^{I}\,\delta A_{I}$, which is independent of $\rho$. But the variations of $X \omega_{\vp }$ and $X^{a}\,A_{\vp a}$ each include terms proportional to $e^{2\rho}$, which must cancel against each other for $\delta \Gamma$ to be defined as $\rho \to \infty$. This is accomplished by setting $b_0 = b_1$. Then we have
\begin{multline}\label{eq:deltaGamma1}
	\delta \Gamma\Big|_\ts{EOM} = \frac{k}{2\pi} \int_{\dM} \extd\varphi \,\bigg[(1 - b_0)\,4\,\mathcal{C}\,\delta \Big(\frac{1}{Y}\Big) + (1 - 2b_0)\,\frac{2}{Y}\,\delta \mathcal{C} \\ 
	+ b_0\, \delta\bigg(L_\ts{CT} - \frac{(\partial_{\varphi}\mathcal{X}^{+})^2}{\mathcal{L}^{+}\mathcal{X}^{+}}\bigg) \bigg] ~,
\end{multline}
where total boundary derivatives have been discarded, and the results have been expressed in terms of $Y$, $\mathcal{C}$, and boundary quantities in the $\mathfrak{sl}(2)$ basis. Requiring the term on the second line to cancel gives a variational principle that fixes, depending on the choice of $b_0$, either $Y$, $\mathcal{C}$, or some combination of these quantities at $\dM$. Since this cancellation only determines the on-shell value of $\delta L_\ts{CT}$ in the $\rho \to \infty$ limit, there are in principle many functions of the fields which are consistent with this condition. A convenient choice which will have a natural interpretation in the second-order formalism is to set
\begin{gather}
    L_\ts{CT} = \frac{(X')^2}{X\,\sqrt{A_{\vp\hat{0}}^{\,2} + A_{\vp\hat{1}}^{\,2}}}~,
\end{gather}
in which case
\begin{gather}
	L_\ts{CT}\Big|_\ts{EOM} = \frac{(\partial_{\varphi}\mathcal{X}^{+})^2}{\mathcal{L}^{+} \mathcal{X}^{+}} +\ldots ~,
\end{gather}
where the ellipsis denotes terms which vanish as $\rho \to \infty$. 

We are primarily interested in the case $b_0 = \tfrac12$, which cancels the $\delta \mathcal{C}$ term in \eqref{eq:deltaGamma1}. This leaves
\eq{
	\delta \Gamma\Big|_\ts{EOM} = \frac{k}{\pi} \int_{\dM} \extd\varphi \,\mathcal{C}\,\delta \Big(\frac{1}{Y}\Big)= \frac{k}{\pi} \,\mathcal{C}\,\int_{\dM} \extd\varphi \,\delta \Big(\frac{1}{Y}\Big) ~,
}{eq:FirstOrderGammaVariation}
which gives a well-defined variational principle with boundary conditions that fix only $1/Y$ on $\dM$. In fact, the boundary conditions are less restrictive than this: since $\mathcal{C}$ is independent of $\varphi$ on-shell, it is sufficient to fix only the zero-mode of $1/Y$. This is the boundary condition we will use in the rest of the paper. Of course, we must also verify that the asymptotic symmetries maintain the value of this zero mode. This will be shown in section \ref{se:4.4}.

Under the gauge transformation \eqref{eq:BoundaryGaugeTransformation}, the bulk term in the action is exactly invariant. The change in our new action $\Gamma$ is a boundary term of the form \eqref{eq:FirstOrderGammaVariation}. And since $1/Y$ transforms as a total derivative on-shell [see \eqref{eq:OtherTransformations}], the constraint $\mathcal{C}'=0$ implies $\delta_{\lambda} \Gamma = 0$. Thus, the action is invariant under the same gauge transformations as the bulk-term \eqref{JT1st}. Evaluated on-shell, this action is non-zero and given by
\begin{equation}
  \label{eq:44}
	\Gamma\,\big|_\ts{EOM} = - \frac{k}{\pi} \,\mathcal{C}\,\int_{\dM} \frac{\extd\varphi}{Y}\,. 
\end{equation}

We have therefore constructed an action for the PSM with the same equations of motion and invariances as the bulk action, but which is non-zero for solutions of the theory. These properties extend to other values of $b_0 = b_1$. In that case, $\delta_{\lambda}\Gamma$ will have the form \eqref{eq:deltaGamma1}, which vanishes by \eqref{eq:OtherTransformations} and the on-shell condition $\mathcal{C}'= 0$. Thus, $b_0$ parameterizes a family of actions with well-defined variational principles, which are invariant under the same gauge transformations as the bulk term \eqref{JT1st} and are generically non-zero (and finite) when evaluated on solutions of the theory. For our choice $b_0=\tfrac12$ the on-shell action \eqref{eq:44} essentially is given by the value of the Casimir ${\cal C}$.

\section{Second order formalism}\label{se:2nd}

The boundary degrees of freedom and symmetries of the JT model are especially clear in the first order formalism, but it is useful to see how these emerge in the more familiar second order formalism. In this section we provide the second order action for the theory, recover the boundary degrees of freedom found in the previous section, and show how the gauge transformations of the PSM arise from bulk diffeomorphisms.

\subsection{Action}
In the second-order formalism, the variables are the dilaton $X$ and the metric $g_{\mu\nu}$ defined on the manifold $\MM$. The action consists of the usual bulk term, a GHY boundary term with the standard coefficient, and a pair of boundary terms (also known as ``holographic counterterms'')
\begin{align}
	\Gamma = &\,\, -\frac{1}{2\kappa^2} \int_{\mathcal{M}} \nts \extd^2x \sqrt{g}\,X\,(R + 2) - \frac{1}{\kappa^2} \int_{\dM} \bns \extd x\,\sqrt{\gamma}\,X\,K \nonumber \\ 
	& \,\, + \frac{1}{\kappa^2} \int_{\dM} \bns \extd x\,\sqrt{\gamma}\,\Big(\sqrt{X^2 + c_0} + \frac{1}{2\,X}\,\gamma^{\mu\nu}\,\partial_{\mu} X \partial_{\nu} X\Big)~,\label{2ndOrderAction}
\end{align}
where $\gamma$ is the induced metric on $\dM$, $K$ is the trace of the extrinsic curvature of $\dM$ embedded in $\mathcal{M}$, and $\kappa^2$ is related to the PSM coupling $k$ by $1/\kappa^{2} = k/(2\pi)$. The variation of this action yields
\begin{align}\label{FullActionVariation}
	\delta \Gamma = &\,\, \frac{k}{4\,\pi} \int_{\cal{M}} \nts \extd^2x \, \sqrt{g}\,\Big[ \, \EE^{\mu\nu}\,\delta g_{\mu\nu} + \EE_{X}\,\delta X \, \Big] \\ \nonumber
		& \,\, + \frac{k}{2\pi} \int_{\dM} \bns \extd x \, \sqrt{\gamma}\,\Big[\, \big(\pi^{\mu\nu} + p^{\mu\nu}\big) \, \delta g_{\mu\nu} + \big(\pi_X + p_{X}\big) \, \delta X\,\Big] ~.
\end{align}
Setting to zero the bulk terms gives the equations of motion
\begin{subequations}
\label{eq:eom}
\begin{align}
	\EE_{\mu\nu} = &\,\, g_{\mu\nu}\,X + \nabla_{\mu} \nabla_{\nu} X - g_{\mu\nu}\,\nabla^{2} X = 0 \\
	\EE_{X} = &\,\, R + 2  = 0~,
\end{align}
\end{subequations}
while the coefficients of the field variations appearing in the boundary term are
\begin{align}
	\pi^{\mu\nu} = &\,\, \frac{1}{2}\,\gamma^{\mu\nu}\,n^{\lambda}\nabla_{\lambda} X \\
	p^{\mu\nu} = &\,\, -\frac{1}{2}\,\gamma^{\mu\nu} \sqrt{X^2 + c_0} + \frac{1}{2\,X}\,\Big(\gamma^{\mu\lambda}\,\gamma^{\nu\sigma} - \frac{1}{2}\,\gamma^{\mu\nu}\,\gamma^{\lambda\sigma}\Big)\,\partial_{\lambda} X \, \partial_{\sigma} X \\
	\pi_{X} = &\,\, K \\
	p_{X} = &\,\, - \frac{X}{\sqrt{X^2 + c_0}} + \frac{1}{2\,X^2}\,\gamma^{\mu\nu}\,\partial_{\mu} X \, \partial_{\nu} X + D_{\mu}\Big( \frac{1}{X}\,D^{\mu} X \Big) ~.
\end{align}
The $\pi$'s come from the variation of the terms in the first line of \eqref{2ndOrderAction}, while the $p$'s come from the variation of the holographic counterterms in the second line.

The first holographic counterterm in the second line of \eqref{2ndOrderAction} was obtained in \cite{Grumiller:2007ju} via variational arguments and the Hamilton--Jacobi approach to holographic renormalization. However, that derivation assumed that the boundary $\dM$ was an isosurface of the dilaton, which is not the case here. Solving the Hamilton--Jacobi equation order-by-order in a boundary derivative expansion yields the final term in \eqref{2ndOrderAction}. The first holographic counterterm contains a constant $c_0$. This constant was set to zero in \cite{Grumiller:2007ju}, to preserve a stringy symmetry of the action (Buscher duality). We will not set it zero immediately. Indeed, we will find in section \ref{sec:schwarz} that it has a natural interpretation in terms of the conformal quantum mechanics of dAFF \cite{deAlfaro:1976je}.

\subsection{Solutions and general boundary conditions}
\label{sec:solbc}

Let us now solve the second order equations of motion \eqref{eq:eom}. To recover the content of the PSM, we partially fix the gauge by setting $g_{\rho\rho} = 1$, and consider metrics which can be written in the form
\begin{gather}
	\extd s^2 = \extd\rho^2 + 2\,j(\vp)\,\extd\rho \,\extd\vp + \Big(\,h(\rho,\vp)^2 + j(\vp)^{2} \,\Big)\,\extd\vp^2 ~.
\end{gather}
The Ricci scalar is then $R = -2\,h^{-1}\,\partial_{\rho}{}^{2}h$, so the equation of motion $\EE_{X} = 0$ immediately gives
\begin{gather}\label{Metric}
	h(\rho,\vp) = e^{\rho}\,\mathcal{L}^{+}(\vp) - e^{-\rho}\,\mathcal{L}^{-}(\vp) ~ .
\end{gather}
Likewise, combining the different components of $\EE_{\mu\nu} = 0$ yields the following equation for the dilaton
\begin{gather}
	\partial_\rho^2 X = X ~,
\end{gather}
which is solved by
\begin{gather}\label{Dilaton}
	X(\rho,\vp) = e^{\rho}\,\mathcal{X}^{+}(\vp) + e^{-\rho}\,\mathcal{X}^{-}(\vp) ~ .
\end{gather}
Comparing with \eqref{AdSLineElement}-\eqref{AdSDilaton}, and making the identification $j = \anot$, the functions appearing in the metric and dilaton are precisely the PSM variables in the $\mathfrak{sl}(2)$ basis.

The contraints of the PSM are obtained from various components of the $\EE_{\mu\nu} = 0$ equations of motion. The equations $\EE_{\rho\vp}=0$ and $\EE_{\rho\rho}=0$ together give the condition
\begin{gather}
	\partial_{\rho}\left(\frac{\anot\,\partial_{\rho} X - \partial_{\vp} X}{h} \right) = 0 ~.
\end{gather}
The quantity in parentheses must be a function of $\vp$, which we identify as $\mathcal{X}^{0}$. As a result, the functions in \eqref{Metric} and \eqref{Dilaton} satisfy
\eq{
	({\cal X}^\pm)' \pm \big(\mathcal{L}^\pm \,\xnot - \mathcal{L}^{0}\,{\cal X}^\pm\big) = 0 \,,
}{dXConstraints}
where a prime indicates a derivative with respect to $\vp$. Finally, if we evaluate $\EE_{\rho\rho} = 0$ using \eqref{Metric}, \eqref{Dilaton}, and \eqref{dXConstraints}, we find one last condition
\begin{gather}\label{hXConstraint}
	(\xnot)' + 2\big(\mathcal{L}^{+} \, \xm - \mathcal{L}^{-}\,\xp\big) = 0 ~.
\end{gather}
The equations \eqref{dXConstraints}-\eqref{hXConstraint} are equivalent to the PSM equations of motion \eqref{eq:psmeom}.
Thus, a general solution of the JT model in the second order formalism contains the same degrees of freedom and constraints as the PSM.

\subsection{On-shell action and on-shell variation}

Solutions of the equations of motion \eqref{eq:eom} have constant negative curvature, $R=-2$, and hence the bulk term in the action \eqref{2ndOrderAction} vanishes. The non-zero contributions come from the boundary terms
\begin{gather}
	\Gamma\,\big|_\ts{EOM} = - \frac{k}{2\pi} \int_{\dM} \bns \extd x\,\sqrt{\gamma}\,\left(X\,K -\sqrt{X^2 + c_0} - \frac{1}{2\,X}\,\gamma^{\mu\nu}\,\partial_{\mu} X \partial_{\nu} X\right)~.
\end{gather}
Taking the boundary $\dM$ as the $\rho_c \to \infty$ limit of the surface $\rho=\rho_c$, the on-shell value is
\begin{gather}\label{OnShell1}
	\Gamma\,\big|_\ts{EOM} = - \frac{k}{2\pi} \int \extd\varphi \,\frac{\mathcal{L}^{+}}{2\,\xp}\,\Big(4\,\mathcal{C} - c_0 \Big) ~.
\end{gather}
Obtaining this result involved integrating-by-parts, imposing the constraints, and dropping total (boundary) derivatives. Except for the term involving $c_0$, this takes the same value as the first order action in section \ref{se:3-2} with $b_0 = \tfrac 12$.

Evaluating the variation of the action \eqref{FullActionVariation} on a solution of the equations of motion, we have
\begin{gather}
	\delta \Gamma\,\Big|_\ts{EOM} = \frac{k}{2\pi}\int \extd\varphi\,\bigg[\frac{1}{4\,\xp \mathcal{L}^{+}}\,\big( 4\,\mathcal{C} + c_0 \big)\,e^{-2\rho}\,\delta g_{\varphi\varphi} - \frac{\mathcal{L}^{+}}{2\,(\xp)^{2}}\,\big( 4\,\mathcal{C} + c_0 \big)\,e^{-\rho}\,\delta X \,\bigg] ~.
\end{gather}
From the powers of $e^{-\rho}$, we see that on-shell $\delta \Gamma$ vanishes for any variations of the fields that grow more slowly than the leading terms in $g_{\varphi\varphi}$ and $X$ as $\rho \to \infty$. This is what one expects for an action that admits a variational principle with Dirichlet boundary conditions on the fields at $\rho \to \infty$. But if we consider variations of the \emph{leading} terms in the fields, this becomes
\begin{gather}\label{LeadingDeltaGamma}
	\delta \Gamma\,\Big|_\ts{EOM} = \frac{k}{2\pi}\int \extd\varphi\,\bigg[\frac{1}{2\,\xp}\,\big( 4\,\mathcal{C} + c_0 \big)\,\delta \mathcal{L}^{+} - \frac{\mathcal{L}^{+}}{2\,(\xp)^{2}}\,\big( 4\,\mathcal{C} + c_0 \big)\,\delta \xp \,\bigg] ~.
\end{gather}
As in section \ref{se:3-1}, we write $\mathcal{L}^{+} = \xp/Y$ and the on-shell variation reduces to
\begin{gather}\label{eq:OnShellVariation}
	\delta \Gamma\,\Big|_\ts{EOM} =  \frac{k}{4\pi}\int \extd\varphi\,(4\,\mathcal{C} + c_0)\,\delta \Big(\frac{1}{Y}\Big) =  \frac{k}{4\pi}\,(4\,\mathcal{C} + c_0)\,\int \extd\varphi\,\delta \Big(\frac{1}{Y}\Big) ~.
\end{gather}
Thus, the on-shell variation of the action is zero even for variations of the leading terms in the fields, provided the zero-mode of the ratio $\mathcal{L}^{+}/\xp=1/Y$ is held fixed. This is the same result found in the the first order formalism in section \ref{se:3-3}.

\subsection{Diffeomorphisms and asymptotic symmetries}\label{se:4.4}

Under a diffeomorphism $x^{\mu} \to x^{\mu} - \xi^{\mu}$, the bulk fields transform with the Lie derivative $\pounds_\xi$ along the vector field $\xi$. 
\eq{
\delta_{\xi} g_{\mu\nu} = \pounds_{\xi} g_{\mu\nu} = \xi^\al\partial_\al g_{\mu\nu} + g_{\mu\al}\partial_\nu\xi^\al + g_{\nu\al}\partial_\mu\xi^\al\qquad\qquad \delta_{\xi} X = \pounds_{\xi} X = \xi^\al\partial_\al X
}{eq:lie}
Diffeomorphisms which act at $\dM$ but leave the action invariant, modulo diffeomorphisms which reduce to the identity at $\dM$, are the asymptotic symmetries of the theory. We show now, with the help of appendix \ref{app:ModifiedBracket}, that these symmetries are precisely the large gauge transformations of the PSM.

In a neighborhood of $\dM$ ($\rho \to \infty$), the most general diffeomorphism that preserves the gauge $g_{\rho\rho} = 1$ and the generalized Fefferman--Graham form of the fields is given by
\begin{subequations}
     \label{eq:Diffeo}
\begin{align}
	\xi^{\vp} = &\,\, \sigma(\vp) + e^{-2\rho}\,\alpha(\vp) + \mathcal{O}(e^{-4\rho}) \\ 
	\xi^{\rho} = &\,\, \lambda(\vp) - \anot(\vp)\,\xi^{\vp} ~.
\end{align}
\end{subequations}
The action of this diffeomorphism on the fields is
\begin{subequations}
\label{DiffeoTransformhp}
\begin{align}
	\delta_{\xi} \ap = & \,\, \lambda\,\ap -\anot\,\sigma\,\ap + (\sigma\,\ap)' \\ \label{DiffeoTransformj}
	\delta_{\xi} \anot = & \,\, \lambda' - 2\,(\ap)^{2}\,\alpha \\ \label{DiffeoTransformhm}
	\delta_{\xi} \am = & \,\, -\lambda\,\am + \anot\,(\sigma\,\am - \alpha\,\ap) + (\sigma\,\am - \alpha\,\ap)' \\ \label{DiffeoTransformXp}
	\delta_{\xi} \xp = & \,\, \lambda\,\xp - \sigma\,\ap\,\xnot \\ \label{DiffeoTransformX0}
	\delta_{\xi} \xnot = & \,\, -2\,\sigma\,\ap\,\xm + 2\,(\sigma\,\am - \alpha\,\ap)\,\xp \\ \label{DiffeoTransformXm}
	\delta_{\xi} \xm = & \,\, -\lambda\,\xm + (\sigma\,\am - \alpha\,\ap)\,\xnot ~.
\end{align}
\end{subequations}

To recover the transformations found in the PSM, we make a field-dependent mapping between the functions in $\xi^{\mu}$ and the parameters of the large gauge transformation \eqref{eq:BoundaryGaugeTransformation}
\begin{gather}\label{Identification}
	\lambda = - \vepsnot \qquad\qquad \sigma = - \frac{\vepsp}{\ap} \qquad\qquad \alpha = \frac{1}{\ap}\,\vepsm - \frac{\am}{(\ap)^{2}}\,\vepsp ~.
\end{gather}
With this identification, \eqref{DiffeoTransformhp} are equivalent to the transformations in section \ref{se:3-1}. Reproducing the symmetry algebra of the PSM is complicated by the field-dependence of the parameters appearing in the diffeomorphism, and requires the introduction of a modified bracket as in \cite{Barnich:2010eb}. This is discussed in detail in Appendix \ref{app:ModifiedBracket}.

Under the diffeomorphism \eqref{eq:Diffeo}, the response of the on-shell action has the form \eqref{eq:OnShellVariation}. As in the PSM, the ratio $\ap / \xp = 1/Y$ transforms on-shell as a total derivative
\begin{equation}
  \label{eq:deltaf}
	\delta_{\xi} \Big(\frac{1}{Y}\Big)\Big|_{\textrm{\tiny EOM}} = \Big(\frac{\sigma}{Y}\Big)' ~.
\end{equation}
This means in particular that the zero mode of $1/Y$ is not changed, which is a non-trivial consistency check of our variational principle. 
Thus, the action \eqref{2ndOrderAction} is invariant under diffeomorphisms that take the form \eqref{eq:Diffeo} in a neighborhood of $\dM$. The asymptotic symmetries are therefore the same as the large gauge transformations of the PSM.

\section{Schwarzian action}
\label{sec:varprinc}
We have shown that the variational principle is well-defined, in both the first and second-order formulation, if the zero-mode of the ratio $\ap/\xp=1/Y$ is fixed. In this section we clarify the interpretation of this variational principle and, provided with these results, show its relation to the Schwarzian action that rose to prominence recently in the context of SYK~\!(-like) models.

\subsection{Comments on the variational principle}

As equation \eqref{eq:deltaf} shows, the quantity $1/Y$ transforms as a total derivative under an infinitesimal change of the boundary coordinate $\varphi\mapsto \varphi+\sigma(\varphi)$. The quantity $Y$ itself transforms as a vector on-shell
\begin{equation}
  \label{eq:11} \delta_\xi Y\big|_{\textrm{\tiny EOM}}=Y'\sigma-\sigma'Y
\end{equation}
under this infinitesimal change of coordinates and is a well-defined, nowhere vanishing vector field on $\partial{\cal M}$ for the following reasons. For consistency, $\ap$ must be a nowhere vanishing positive function such that the induced metric on the cut-off surface $\rho_c$ is Euclidean and non-singular in the limit $\rho_c\rightarrow \infty$. Similarly, the leading order component of the dilaton, $\xp$, must be a non-zero (positive) function everywhere if we want to interpret the asymptotic region $\rho \rightarrow \infty$ as a weak coupling region $X \rightarrow \infty$. Consequently, the quantity
\begin{equation}
  \label{eq:18}
  \overline{Y}^{-1}\equiv \frac{1}{\beta}\int\limits_0^\beta \frac{\extd \varphi}{Y}
\end{equation}
that we keep fixed as part of our boundary conditions is well-defined.

Furthermore, let us define the mass function $M(\vp)$,
 \begin{equation}
      \label{eq:12} M=\mathcal{T}-\mathcal{P}^2-{\cal P}' 
    \end{equation}
    where
    \begin{equation}
      \label{eq:JandT}
      \mathcal{T}=\mathcal{L}^+\mathcal{L}^- \qquad\qquad  \mathcal{P}=\tfrac12\,\mathcal{L}^0 - \frac{(\mathcal{L}^+)'}{2\mathcal{L}^+}\,.
    \end{equation}
    This can be regarded as a boundary stress tensor obtained by a (twisted) Sugawara construction \eqref{eq:12} from the $\mathfrak{sl}(2)$ generators ${\cal L}^\pm,\mathcal{L}^0$. It transforms with an infinitesimal Schwarzian derivative,
    \begin{equation}
      \label{eq:13}
      \delta_{\xi}{M}=\sigma {M}'+2\sigma'{M}+\frac{1}{2}\sigma'''
    \end{equation}
    under infinitesimal reparametrizations of the boundary coordinate. Under finite transformations, $\varphi \mapsto f(\varphi)$, where $f(\varphi)$ is a diffeomorphism on $S^1$ obeying
    \begin{equation}
      \label{eq:19}
      f'(\varphi)>0\qquad f(\varphi+\beta)=f(\varphi)+\beta
    \end{equation}
    we find the transformation law ${ M}\mapsto \tilde{{ M}}$
    \begin{equation}
      \label{eq:20}
      \tilde{{M}}(f(\varphi))=\frac{1}{(f'(\varphi))^2}\big({ M}(\varphi)-\tfrac{1}{2}\, \textrm{Sch}[f](\varphi)\big).
    \end{equation}
    Here, $\textrm{Sch}[f](\varphi)$ denotes the Schwarzian derivative
    \begin{equation}
      \label{eq:23} \textrm{Sch}[f](\varphi)=\left(\frac{f''}{f'}\right)'-\frac{1}{2}\left(\frac{f''}{f'}\right)^2.
    \end{equation}
    
    The quantity $M$ can therefore be regarded as an element of a specific coadjoint orbit of the Virasoro group \cite{Witten:1987ty}; for a thorough pedagogic treatise with applications to three-dimensional gravity consult \cite{Oblak:2016eij}. 
    In the following it will be convenient to evaluate the left hand side of \eqref{eq:20} at $\varphi$ instead of $f(\varphi)$. Using the inversion formula for the Schwarzian derivative,
    $\textrm{Sch}[f](\varphi)=-\big(f'(\varphi)\big)^2\,\textrm{Sch}[f^{-1}](\varphi)$,
    yields
    \begin{equation}
    \label{eq:20inv}
    \tilde{{M}}(\varphi)=\big((f^{-1})'(\varphi)\big)^2 M(f^{-1}(\varphi))+\tfrac{1}{2}\, \textrm{Sch}[f^{-1}](\varphi)\,.
    \end{equation}
    Since a particular coadjoint orbit is a homogeneous space for the Virasoro group, the result \eqref{eq:20inv} shows that any point on the orbit $\tilde{M}$ can be reached by acting with an appropriate diffeormophism $f(\varphi)$ on a chosen representative $M$. 
    With the help of the quantity $M$, the constraints \eqref{eq:34}-\eqref{eq:34b} are equivalent to the equation
    \begin{equation}
      \label{eq:17}
      {\cal C}=Y^2{ M}-\tfrac{1}{4}(Y')^2+\tfrac{1}{2}Y Y''\,,
    \end{equation}
    relating  $M$, $Y$, and the Casimir function ${\cal C}$. Conservation of the Casimir ${\cal C}^\prime=0$ establishes
    \begin{equation}
      \label{eq:14}
      Y { M}'+2Y'{ M}+\tfrac{1}{2}Y'''=0\,.
    \end{equation}
    We stress that only two of the three constraints are needed to derive equation \eqref{eq:17}, which implies that this equation is valid without assuming the conservation of the Casimir. By contrast, \eqref{eq:14} is an immediate consequence of this conservation and is valid only if all three constraints are imposed. This distinction between ``fully on-shell" and ``partially on-shell" will be important for the discussion in section \ref{se:2.3} and \ref{se:2.2}. 
    
    Since the (rescaled) leading order of the dilaton field transforms like a boundary vector and solves equation \eqref{eq:14} it can be regarded as the \emph{stabilizer} of the coadjoint orbit of the Virasoro group determined by ${ M}$.\footnote{The classification of coadjoint orbits essentially boils down to determine the stabilizers of each orbit. More specifically, if $G$ is the stabilizer group for a coadjoint orbit of the Virasoro group, the respective orbit is given by $\textrm{Diff}(S^1)/G$.} If the on-shell condition of conservation of the Casimir function is not enforced, comparison between \eqref{eq:14} and \eqref{eq:13} suggests that the quantity $Y$ generates infinitesimal diffeomorphisms under which $M$ transforms anomalously. In section \ref{se:2.3} we will see that, with a caveat, this is indeed the case.

    By solving the equation
    \begin{equation}
      \label{eq:24}
      \frac{\extd \varphi}{Y}=\frac{\extd \tilde{\varphi}}{\overline{Y}}
    \end{equation}
    one can always find a diffeomorphism $\varphi\mapsto \tilde{\varphi}$ to a new coordinate system $\tilde{\varphi}$ in which $Y$ takes the constant value $\overline{Y}$. In this coordinate system equation \eqref{eq:17} yields
    \begin{equation}
      \label{eq:25}
      {M}={\cal C}\overline{Y}^{-2},
    \end{equation}
    thus determining the constant representative of each orbit since the Casimir is conserved.\footnote{The full classification of orbits of the Virasoro group includes, in addition to the orbits with constant representative constructed above, an infinite number of families without constant representative. Our conditions on the dilaton field, in particular the requirement that it is non-zero everywhere, disallow these orbits.}
    In this coordinate system the solution of equation \eqref{eq:14} is straightforward. For generic values of $M$, $\overline{Y}$ will be the only periodic solution to this equation, and the stabilizer group is just $U(1)$. However, at the exceptional values
    \begin{equation}
      \label{eq:26}
      { M}=\frac{n^2\pi^2}{\beta^2}
    \end{equation}
    one finds two additional solutions, and the stabilizer group is given by $\textrm{PSL}^{(n)}(2,\mathbb{R})$, i.e., the $n$-fold cover of the Euclidean $\textrm{AdS}_2$ group $\textrm{SO}(2,1)\simeq\textrm{SL}(2,\mathbb{R})/\mathbb{Z}_2$.
    The smooth classical solutions compatible with the choice of temperature are determined by calculating a holonomy around the $\varphi$-cycle and demanding that it equals minus unity,
    \begin{equation}
      \label{eq:37}
      \mathcal{P}\exp\bigg(\oint A\bigg)=-1
    \end{equation}
    where $\mathcal{P}$ denotes path ordering.
    This singles out the Euclidean black hole configurations, i.e., for any choice of inverse temperature $\beta$ these are the constant representative solutions \eqref{eq:26} with $n=1$. The relation between Casimir and temperature for smooth classical solutions is therefore given by
    \begin{equation}
      \label{eq:38}
      {\cal C}=\frac{\bar{Y}^2\pi^2}{\beta^2}.
    \end{equation}
The fact that the Casimir $\cal C$ scales quadratically with temperature $T=1/\beta$ is compatible with the two-dimensional Stefan--Boltzmann law.

\subsection{Schwarzian action}\label{se:black}
\label{sec:schwarz}
In this section we make contact with the recent developments regarding a proposed duality between (nearly) $\textrm{AdS}_2$ gravity in the form of the JT model and the SYK model \cite{Sachdev:1992fk, Kitaev:15ur}. This quantum mechanical model of Majorana fermions with a four-point interaction with random coupling develops a conformal symmetry in the strong coupling/low energy regime, i.e., it allows for arbitrary reparametrizations of (Euclidean) time. This symmetry, however, is spontaneously broken to an $\textrm{SL}(2)$ symmetry by the groundstate. The low energy dynamics of the theory is therefore governed by the reparametrizations that become Nambu--Goldstone bosons due to this spontaneous symmetry breaking and acquire an effective action given by the Schwarzian action \cite{Maldacena:2016hyu}. The Schwarzian action provides the link to $\textrm{AdS}_2$ gravity as it was shown that the effective dynamics of the JT model can be rewritten also in the form of a Schwarzian action \cite{Maldacena:2016upp,Jensen:2016pah,Engelsoy:2016xyb}. In the following we will show that our on-shell action \eqref{OnShell1}, deriving from an action with well-defined variational principle both in the first \eqref{eq:JT1} and second order formulation \eqref{2ndOrderAction}, can be naturally reformulated as a Schwarzian action.

Using the notation introduced in section~\ref{sec:varprinc}, the on-shell action takes the form
\begin{equation}
  \label{eq:effective1}
  \Gamma\,\big|_\ts{EOM} = - \frac{k}{4\pi} \int\limits_0^\beta \frac{\extd\varphi}{Y}\,\Big(4\,\mathcal{C} - c_0 \Big)=-\frac{k}{4\pi}\int\limits_0^{\beta} \frac{\extd\varphi}{Y}\left(4Y^2{ M}-(Y')^2-c_0\right),
\end{equation}
where we used equation \eqref{eq:17} in the second step and discarded a total derivative. The mass function $M$ must be an element of the Virasoro orbit with constant representative given by \eqref{eq:26} with $n=1$ since otherwise we would have a solution that is not smooth for given $\beta$.

As a first observation note that \eqref{eq:effective1} becomes the action of Euclidean conformal quantum mechanics discussed in \cite{deAlfaro:1976je,Astorino:2002bj} coupled to the external source $M$ upon replacing $Y\rightarrow q^2$.
\begin{equation}
  \label{eq:40}
  \Gamma = - \frac{k}{\pi} \int\limits_0^\beta \extd\varphi\left(q^2 M-(q')^2-\frac{c_0}{4q^2}\right)
\end{equation}
As mentioned above, the quantity $c_0$ becomes the coupling strength of the conformal quantum mechanics model. Consistent with $Y$ transforming like a boundary vector under arbitrary reparametrizations $q$ should transform with conformal weight $-\frac{1}{2}$.

We return now to \eqref{eq:effective1} and set $c_0=0$. This value is special since for string-related models of dilaton gravity it restores a stringy symmetry, Buscher duality \cite{Buscher:1987sk}, while for JT it restores homogeneity of the action in the dilaton field $X$.
Let us define a diffeomorphism $g:S^1\rightarrow S^1,\varphi\mapsto u=g(\varphi)$ by
\begin{equation}
  \label{eq:36}
g(\varphi)= \bar{Y}\int\limits_0^\varphi\frac{\extd \eta}{Y(\eta)}\,.
\end{equation}

This is a finite reparametrization of the boundary coordinate $\varphi$. We can therefore rewrite the action \eqref{eq:effective1} as
\begin{equation}
  \label{eq:41}
\Gamma=-\frac{k \bar{Y}}{\pi}\int\limits_0^\beta \extd u\, \Big( (g^{-1})'(u)M +\tfrac{1}{2}\,\textrm{Sch}[g^{-1}](u)\Big)\,.
\end{equation}
The Lagrangian in \eqref{eq:41} is the coadjoint action of the Virasoro group \eqref{eq:20inv} acting on the element $M$ and provides an effective action for reparametrizations $g^{-1}(u)$. One should not consider independent variations of $M$ and $g^{-1}(u)$ when varying \eqref{eq:41} but rather impose the constraint \eqref{eq:14} on the variations. Furthermore, variations of $M$ must not leave the orbit of the constant representative that is consistent with the choice of temperature $T=\beta^{-1}$.

Without loss of generality we assume $M$ is a constant representative (since any element on the orbit can be reached from it), and setting $g^{-1}(u)\equiv \tau(u)$ we find
\begin{equation}
  \label{eq:43}
  \Gamma=-\frac{k \bar{Y}}{2\pi}\int\limits_0^\beta \extd u\,\bigg(\frac{1}{2}\Big(\frac{2\pi}{\beta}\Big)^2(\tau')^2+\textrm{Sch}[\tau](u)\bigg)\,,
\end{equation}
which is precisely the Schwarzian action at finite temperature $\beta$ for finite reparametrizations of the circle $\tau$ \cite{Maldacena:2016hyu, Maldacena:2016upp}.

\section{Menagerie of AdS\texorpdfstring{$_2$}{2} boundary conditions}\label{se:2}

Having specified our variational principle we now turn to the calculation of the asymptotic charges.

The canonical boundary current for dilaton gravity in the PSM formulation can be obtained using covariant \cite{Wald:1999wa, Barnich:2001jy} or canonical approaches \cite{Regge:1974zd}. Both yield the expression
\eq{
\delta Q[\varepsilon] = \frac{k}{2\pi}\,\varepsilon_{I}\,\delta X^{I}=\frac{k}{\pi}\,\tr\left(\varepsilon\,\delta \mX\right)\,.
}{eq:deQ}
  The currents \eqref{eq:deQ} are to be evaluated at the asymptotic boundary of one Euclidean time-slice, i.e., they are valid at one particular value of angular coordinate $\varphi$. However as suggested in \cite{Cadoni:2000ah}, we will define time-averaged versions of the canonical boundary currents as
  \begin{equation}
    \label{eq:28}
    \delta \tilde{Q}[\varepsilon] = \frac{k}{2\pi\beta}\int\limits_0^\beta \extd \varphi\, \varepsilon_{I}\,\delta X^{I}=\frac{k}{\pi\beta}\int\limits_0^\beta \extd \varphi\, \tr\left(\varepsilon\,\delta \mX\right).
  \end{equation}
As stated in the introduction, these time-averaged boundary currents depend on the full tower of Fourier modes of the transformation parameters $\eps_I$ and field variations $\delta X^{I}$. 

In the next subsections we specify four different sets of boundary conditions (with a fifth one in appendix \ref{app:B}), integrate the time-averaged boundary currents \eqref{eq:28} in field space to averaged charges and study their associated algebras.
  
\subsection{Loop group boundary conditions}\label{se:2.1}

We start with the loosest set of boundary conditions where no restrictions are placed on $\delta\mX$ or $\delta{\cal L}^I$ other than on-shell conditions [and the fixing of the zero mode of $1/Y$ as defined in \eqref{eq:Y}]. This means that the metric and dilaton have to obey the boundary conditions
\eq{
g_{\mu\nu}\,\extd x^\mu\extd x^\nu = \extd\rho^2 + {\cal O}(1)\, \extd\rho\extd\vp + \big({\cal O}(e^{2\rho}) + {\cal O}(1) + \dots\big)\,\extd\vp^2\qquad X = {\cal O}(e^\rho) + {\cal O}(e^{-\rho})  
}{eq:angelinajolie}
where the ellipsis refers to a term of order ${\cal O}(e^{-2\rho})$ that is fully determined by on-shell conditions. The leading order variations of $g_{\vp\vp}$ and the dilaton are subject to the condition that the ratio $\sqrt{g_{\vp\vp}}/X$ has a zero mode that is not allowed to vary, as explained in sections \ref{se:3} and \ref{se:2nd}.

Assuming that the transformation parameters $\eps_I$ in \eqref{eq:28} are field-independent allows integration in field space and leads to the averaged charges
\begin{equation}
  \label{eq:charge}
\tilde{Q}[\varepsilon]	=	\frac{k}{\pi\beta}\int\limits_0^\beta \extd\varphi\:\left[\frac{1}{2}\varepsilon^{0}\mathcal{X}^{0}-\varepsilon^{+}\mathcal{X}^{-}-\varepsilon^{-}\mathcal{X}^{+}\right]\,.
\end{equation}
In the following we will always refer to the quantities defined in \eqref{eq:charge} simply as ``charges'' and drop the tilde for notational brevity.

Using the fact that the charges \eqref{eq:charge} generate symmetry transformations according to
\begin{equation}
  \label{eq:29}
  \delta_\varepsilon F=\{F,Q\left(\varepsilon\right)\}\, ,
\end{equation}
we can determine the brackets between elements of the asymptotic phase space spanned by ${\cal X^\pm},{\cal X^0}$ and $\apm,\anot$.
In particular, defining the Fourier coefficients of the rescaled charges
\begin{equation}
  \label{eq:27}
  \mathcal{X}^{\pm}_n=\frac{1}{\beta}\int\limits_0^\beta\dd \varphi\, e^{in\varphi}{\cal X}^\pm\qquad \mathcal{X}^{0}_n=\frac{1}{2\beta}\int\limits_0^\beta\dd \varphi\, e^{in\varphi}{\cal X}^0,
\end{equation}
we find that their algebra is given by a centerless $\widehat{\mathfrak{sl}}(2)$ current algebra
\begin{equation}
  \left\{ \mathcal{X}_{n}^{I},\mathcal{X}_{m}^{J}\right\} = (I-J)	\mathcal{X}_{n+m}^{I+J}\,.
\end{equation}
This agrees with the algebra of asymptotic Killing vectors equipped with the modified Lie bracket of \cite{Barnich:2010eb} in the second order formalism, as explained in more detail in appendix \ref{app:ModifiedBracket}, and also with the asymptotic symmetries in the PSM formulation \eqref{eq:deX}.

\subsection{Conformal boundary conditions}\label{se:2.3}

We will now turn to the set of stricter boundary conditions that was analyzed previously in \cite{Grumiller:2013swa, Grumiller:2015vaa} and that is more ``typical'' for asymptotically AdS$_2$ behavior since the leading order metric is not allowed to fluctuate.
Setting $\anot=0$ and fixing $\ap$ to the convenient constant value $\ap=1/2$, metric and dilaton read
\eq{
g_{\mu\nu}\,\extd x^\mu\extd x^\nu = \extd\rho^2 + \big(\tfrac14\,e^{2\rho} + {\cal O}(1) + \dots\big)\,\extd\vp^2\qquad X = {\cal O}(e^\rho) + {\cal O}(e^{-\rho})  
}{eq:lalapetz}
where the ellipsis refers to a term of order ${\cal O}(e^{-2\rho})$ that is fully determined by on-shell conditions. The leading order variation of the inverse of the dilaton is subject to the condition that its zero mode is not allowed to vary (for the same reasons as above).

The auxiliary connection $a$ for these boundary conditions is given by
\begin{equation}
  \label{eq:concon}
 a = \frac{1}{2}L_+  +  {\cal L}^-(\vp) L_-\,.
\end{equation}
Let us define $\sigma\equiv-\varepsilon^{+}/\ap=-2\varepsilon^+$, as in \eqref{Identification}, and $\mathcal{L}\equiv\mathcal{L}^{-}$. Then the conditions $\delta\mathcal{L}^0=\delta\mathcal{L}^+=0$ imply the following relations between the gauge parameters:
\begin{align}
  \label{eq:con0}
  \varepsilon^{0}&=\partial_{\varphi}\sigma={\sigma}^\prime\\
  \label{eq:con-}
\varepsilon^{-}	&=-{\sigma}''-\sigma\mathcal{L}\,.
\end{align}
The function $\mathcal L$ transforms with an infinitesimal Schwarzian derivative
\begin{equation}
\delta_{\sigma}\mathcal{L}=\sigma{\mathcal{L}^\prime}+2{\sigma}^\prime\mathcal{L}+{\sigma}'''\label{cbc01}
\end{equation}
and is related to the mass function by a factor $\tfrac12$
\eq{
{M}=\frac{1}{2}\,\am=\frac12\,\mathcal{L} \,.
}{eq:ML}
Therefore, also the mass function $M$ again transforms with an infinitesimal Schwarzian derivative [as in \eqref{eq:13}] under infinitesimal diffeomorphisms parametrized by $\sigma$. 

The asymptotic symmetries for these boundary conditions were previously analyzed in \cite{Grumiller:2013swa, Grumiller:2015vaa}. It was shown therein that the charges associated with these asymptotic symmetries are, in general, non-integrable. In the present context one might be tempted to insert the parameters \eqref{eq:con0} and \eqref{eq:con-} into \eqref{eq:charge} and declare the result to be the asymptotic symmetry generators that canonically realize the Virasoro symmetry apparent in \eqref{cbc01}. However, the asymptotic charges equipped with the Poisson bracket \eqref{eq:29} do not form an algebra; rather one would have to construct the Dirac brackets implementing the constraints we imposed to arrive at \eqref{eq:concon}. We shall not construct these brackets in the present paper but will follow a different way to deal with the non-integrability of the charges.\footnote{A prescription to calculate the Poisson brackets of non-integrable charges was presented in \cite{Barnich:2011mi}. The price one has to pay in that approach is a non-standard central extension, in the sense that it becomes field-dependent.}

The variation of the time-averaged charges \eqref{eq:28} takes the form
\begin{equation}
\label{eq:confch}
\delta Q[\sigma]=\frac{k}{\pi \beta}\int\limits_0^\beta\extd \varphi \left(\frac{1}{2}{\sigma}^\prime\delta \xnot+\frac{1}{2}\sigma \delta \xm+\frac{1}{2}{\sigma}''\, \delta Y+\frac{1}{2}\sigma {\cal L} \delta Y\right)\,,
\end{equation}
where we used relations \eqref{eq:con0} and \eqref{eq:con-} for the gauge parameters and defined again $Y=\xp/\ap=2\xp$. It is obvious that the last term of this expression spoils integrability. Using the linearized equations of motion, the variations of \eqref{eq:34} and \eqref{eq:34b}, to eliminate $\delta {\mathcal X}^-$ and $\delta{\mathcal X}^0$ the charge \eqref{eq:confch} can be rewritten as
\begin{equation}
  \label{eq:bupi}
\delta Q[\sigma]=\frac{k}{2 \pi \beta}\int\limits_0^\beta\extd \varphi \left(-{\sigma}^\prime\delta {Y}^\prime+\sigma \delta {Y}''+\sigma \delta ({\cal L} Y)+{\sigma}''\, \delta Y+\sigma {\cal L} \delta Y\right)\,,
\end{equation}

The expression \eqref{eq:bupi} is still non-integrable. However, in section \ref{sec:varprinc} we saw that the quantity $Y$ is a boundary vector that is related to infinitesimal reparametrizations of the boundary coordinate as suggested by equation \eqref{eq:14}. We therefore redefine the gauge parameter $\sigma$ as 
\eq{
\sigma=\varepsilon Y\qquad  \textrm{with}\quad \partial_{\varphi}\varepsilon=0=\delta\eps
}{eq:sieps}
The redefinition \eqref{eq:sieps} effectively amounts to a change of our boundary conditions. As we show now it leads to integrable charges with interesting properties. Inserting the redefinition \eqref{eq:sieps} into the variation of the charges \eqref{eq:bupi} we find that the charges become integrable.
\begin{equation}
  \label{eq:30}
  Q[\sigma]=\frac{k}{\pi\beta}\int\limits_0^\beta \extd \varphi\, \frac{\sigma}{Y} \bigg(Y^2M -\frac{1}{4} {Y^\prime}^2+\frac{1}{2}Y {Y}''\bigg)
\end{equation}

The quantity in parentheses is just the Casimir \eqref{eq:17}. However, let us not enforce the on-shell conservation of the Casimir for the moment. Then, following the same line of reasoning that led us to the Schwarzian action in section \ref{sec:schwarz}, we find that the charge \eqref{eq:30} is given by
\begin{equation}
  \label{eq:33}
  Q[\sigma]=\frac{k\bar{Y}}{2\pi\beta}\int\limits_0^\beta \extd u\,\sigma(u) \left(\frac{1}{2}\left(\frac{2\pi}{\beta}\right)^2(\tau')^2+\textrm{Sch}[\tau|(u)\right)\,.
\end{equation}
By equation \eqref{eq:20inv} the quantity in parentheses denotes a generic point $M(u)$ on the orbit of the constant representative $\textrm{Diff}(S^1)/\textrm{SL}(2,\mathbb{R})$, which leads to
\begin{equation}
  \label{eq:35}
   Q[\sigma]=\frac{k\bar{Y}}{\pi\beta}\int\limits_0^\beta \extd u\,\sigma(u) M(u)\,.
\end{equation}
These are just the usual charges on would expect on the phase space of a coadjoint orbit of the Virasoro group \cite{Witten:1987ty}. Indeed, using equation \eqref{eq:29} one can check that
\begin{equation}
  \label{eq:42}
  \left\{Q[\sigma_1],Q[\sigma_2]\right\}=\frac{k\bar{Y}}{\pi\beta}\int\limits_0^\beta \extd u\,\sigma_1(u) \left(\sigma_2M' + 2\sigma_2'M + \tfrac{1}{2}\sigma_2'''\right)
\end{equation}
since $\delta_{\sigma_2}\sigma_1=0$ due to the relations \eqref{eq:11} and \eqref{eq:sieps}. We therefore find a Virasoro algebra at central charge
\begin{equation}
c=\frac{6k\bar{Y}}{\pi}\label{eq:43b}
\end{equation}
where $\bar Y$ is defined in \eqref{eq:18}. Thus, our requirement of a well-defined variational principle that led to the fixing of $\bar Y$ implies that the central charge \eqref{eq:43b} is state-independent.

However, the above derivation was based on the assumption that the strict on-shell conservation of the Casimir is not enforced. This is also clear from equation \eqref{eq:42} since recalling the parametrization $\sigma_2=\varepsilon_2 Y$ we obtain
\begin{equation}
  \label{eq:43a}
  \left\{Q[\sigma_1],Q[\sigma_2]\right\}=\frac{k\bar{Y}}{\pi\beta}\int\limits_0^\beta \extd u\,\sigma_1(u)\varepsilon_2 \left(2Y'M+Y M'+\frac{1}{2}Y'''\right)=0,
\end{equation}
due to equation \eqref{eq:14} which was a consequence of the on-shell conservation of the Casimir.

In summary, the above discussion suggests the following general picture: Using the conformal boundary conditions \eqref{eq:concon} we find that off-shell the averaged charges equipped with the Poisson brackets \eqref{eq:29} form a Virasoro algebra. However, the on-shell conservation law of the Casimir breaks this conformal symmetry to a simple $U(1)$ with the generator given by the $\textrm{SL}(2,\mathbb{R})$-invariant Casimir. 

This pattern of on-shell breaking of conformal symmetry is a distinctive feature of the SYK model \cite{Maldacena:2016upp,Maldacena:2016hyu, Engelsoy:2016xyb, Jensen:2016pah, Kitaev:15ur}.

\subsection{Warped conformal boundary conditions}\label{se:2.2}

A looser set of boundary conditions than conformal ones may be obtained if one sets to zero the $L_0$ part of the connection but does not fix the leading order term in the metric,
\begin{equation}
    \mathcal{L}_0=0\,.\label{eq:L00}
\end{equation}
The corresponding boundary conditions on metric and dilaton read
\eq{
g_{\mu\nu}\,\extd x^\mu\extd x^\nu = \extd\rho^2 + \big({\cal O}(e^{2\rho}) + {\cal O}(1) + \dots\big)\,\extd\vp^2\qquad X = {\cal O}(e^\rho) + {\cal O}(e^{-\rho})  
}{eq:whatever}
where the ellipsis refers to a term of order ${\cal O}(e^{-2\rho})$ that is fully determined by on-shell conditions. The leading order variations of $g_{\vp\vp}$ and the dilaton are subject to the condition that the ratio $\sqrt{g_{\vp\vp}}/X$ has a zero mode that is not allowed to vary (for the same reasons as above).

The condition $\delta_\varepsilon \mathcal{L}^0=0$ gives a restriction on the parameters of large gauge transformations
\begin{equation}\label{WCepsmCond}
\partial_\varphi \varepsilon^0=2\varepsilon^+\mathcal{L}^- -2\varepsilon^-\mathcal{L}^+ \, ,
\end{equation}
while $\mathcal{L}^\pm$ transform as follows
\begin{eqnarray}
&&\delta_\varepsilon \mathcal{L}^+=-\partial_\vp\varepsilon^+ -\varepsilon^0\mathcal{L}^+ \\
&&\delta_\varepsilon \mathcal{L}^-=-\partial_\vp\varepsilon^- +\varepsilon^0\mathcal{L}^-.
\end{eqnarray}
Rescaling the gauge parameter $\varepsilon^+$ by defining $\sigma\equiv-\varepsilon^+/\ap$, which corresponds to an infinitesimal reparametrization $\varphi\mapsto \varphi+\sigma$, and setting $\varepsilon^0=-\lambda$  according to \eqref{Identification} we find that the quantities $\mathcal{P}$ and $\mathcal{T}$ defined in \eqref{eq:JandT} transform as
\begin{subequations}
    \label{eq:warpedcoad}
\begin{align}
\delta_\varepsilon \mathcal{P} &= -\frac{1}{2}{\sigma}''+{\sigma}^\prime\mathcal{P}+\sigma{\mathcal{P}}^\prime-\frac{1}{2}{\lambda}^{\prime}\\
\delta_\varepsilon \mathcal{T} &= \sigma{\mathcal{T}}^\prime+2{\sigma}^\prime\mathcal{T}-{\lambda}^{\prime}\mathcal{P}-\frac{1}{2}{\lambda}^{\prime\prime}\,.
\end{align}
\end{subequations}
Notice that $\mathcal{P}$ reduces to $\mathcal{P}=-\frac{1}{2}\partial_{\varphi}\log \ap$ due to \eqref{eq:L00}.
This transformation behaviour is characteristic of a warped conformal algebra \cite{Detournay:2012pc} with twist term \cite{Afshar:2015wjm}. Our boundary conditions can thus be regarded as a two-dimensional analog of the $\textrm{AdS}_3$ boundary conditions introduced in \cite{Troessaert:2013fma}. In the present case, the quantity $M$ defined in \eqref{eq:12} still transforms anomalously
\begin{equation}
\delta_{\varepsilon} {M}	=	\sigma{{M}}^\prime+2{\sigma}^\prime{M}+\frac{1}{2}{\sigma}'''\,.\label{wcbc02}
\end{equation}
To get rid of the twist term in the transformations \eqref{eq:warpedcoad} one can redefine the generators
\eq{
\hat M:=M+\alpha{\hat P}^2\qquad\alpha\in\mathbb{R}^+\qquad\qquad\hat P:=\mathcal{P}-\tfrac12\,\partial_\vp\ln Y
}{eq:wcft5}
which transform as
\begin{subequations}
    \label{eq:nicewcft}
\begin{align}
\delta_\varepsilon \hat P &= {\sigma}^\prime\hat{P}+\sigma{\hat{P}}^\prime-\frac{1}{2}{\lambda}^{\prime}\\
\delta_\varepsilon \hat M &= \sigma{\hat{M}}^\prime+2{\sigma}^\prime\hat{M}+\frac12\,\si'''-\alpha {\lambda}^{\prime}\hat{P}\,.
\end{align}
\end{subequations}

We construct the time-averaged charges using the same approach as the previous section. Starting from the expression \eqref{eq:28} for $\delta Q$, we use the condition \eqref{WCepsmCond} to replace $\vepsm$ and the linearized equation of motion \eqref{eq:34b} to rewrite $\am$. Then, with $\xp = \ap\,Y$ as before, $\delta Q$ is
\begin{align}
    \delta Q = \frac{k}{\pi\,\beta} \int\limits_{0}^{\beta}\extd\varphi\,\bigg[\frac12\,\vepsnot\,\delta\xnot-\vepsp\,\delta\xm-\frac{\xm}{\xp}\vepsp\,\delta\xp + \frac{Y}{2(\xp)^2}\big(\xp(\vepsnot)^\prime-(\xnot)^\prime\vepsp\big)\delta\xp\bigg]\,.
\end{align}
This is rendered integrable by the following redefinition of the gauge parameters
\begin{gather}
\label{eq:wcred}
    \vepsp = \veps\,\xp \qquad \qquad \vepsnot = \veps\,\xnot + \eta 
\end{gather}
with ${\veps}^\prime = {\eta}^\prime = 0$. The time-averaged charge is then
\begin{gather}
    Q = \frac{k}{\pi\,\beta}\,\int\limits_{0}^{\beta} \extd\varphi \, \bigg[ \veps \Big(\frac{1}{4}\,(\xnot)^{2} - \xp\,\xm\Big) + \eta\,\frac{1}{2}\,\xnot \bigg] ~.
    \label{eq:wcft0}
\end{gather}
The quantity in parentheses is $-\mathcal{C}$ [cf.~the definition of the Casimir \eqref{eq:Casimir}]. The term proportional to $\eta$ can be expressed in terms of $\cal P$, so that \eqref{eq:wcft0} can be rewritten as
\eq{
 Q = \frac{k}{\pi\,\beta}\,\int\limits_{0}^{\beta} \extd\varphi \, \bigg[ -\veps \Big(Y^2M-\frac14\,(Y')^2+\frac12\,YY''\Big) + \eta\,\Big(Y{\cal P} - \frac12\,Y'\Big) \bigg] ~.
}{eq:wcft1}
Going fully on-shell the result \eqref{eq:wcft1} reduces to the sum of two zero-mode charges.
\eq{
 Q\big|_{\textrm{\tiny EOM}} = -\eps\,\frac{k}{\pi}\, {\cal C} + \eta\,\frac{k}{\pi\beta}\int\limits_{0}^{\beta} \extd\varphi \, P_0 
}{eq:wcft2}
where 
\eq{
P_0:=-\tfrac12\,Y\partial_\vp\ln X^+ = Y\hat P =\tfrac12\,X^0\,. 
}{eq:warped0}
The fact that there are two zero mode charges is in agreement with the warped conformal interpretation of our boundary conditions. 

Expressing the charges off-shell (by analogy to section \ref{se:2.3}) in terms of $\si=-\eps^+/\ap$ and $\la=-\eps^0$ using \eqref{eq:wcred} yields
\eq{
 Q = \frac{k}{\pi\,\beta}\,\int\limits_{0}^{\beta} \extd\varphi \, \bigg[ \frac{\si}{Y}\, \Big(Y^2\big(M+2{\hat{P}}^2\big)-\frac14\,(Y')^2+\frac12\,YY''\Big) - \la\,Y\hat{P} \bigg] ~.
}{eq:wcft3}
It is gratifying that the result \eqref{eq:wcft3} contains the redefined quantities \eqref{eq:wcft5} [with $\alpha=2$] that transform like a Virasoro and a $\mathfrak{u}(1)$ current algebra with no twist term \eqref{eq:nicewcft}.\footnote{However, unlike in section \ref{se:2.3} we were not able to reproduce \eqref{eq:nicewcft} directly from varying \eqref{eq:wcft3}.}

\subsection{\texorpdfstring{$u(1)$}{u(1)} boundary conditions}\label{se:2.4}

Another case of interest is to keep the $L_0$ component in $a$ arbitrary and to fix $\mathcal L^\pm =0$. 
\begin{equation}
 a = L_0 {\cal L}^0(\vp) \,\extd\vp
\end{equation}
We still have a well-defined variational principle in this case since \eqref{LeadingDeltaGamma} vanishes identically for $\ap=0$. The variations \eqref{eq:lgt} are compatible with these choices if we fix two of the variation parameters to zero, $\eps^\pm=0$. 
\begin{align}
\de_\eps {\mathcal L}^\pm &= 0 \\
\de_\eps {\mathcal L}^0 &=-\eps^{0\,\prime} \\
\de_\eps {\mathcal X}^\pm &= \mp \eps^0{\mathcal X}^\pm\\
\de_\eps {\mathcal X}^0 &= 0
\end{align}
Notably, the metric function ${\mathcal L}_0$ transforms like a $\mathfrak{u}(1)_k$ current algebra and the only non-trivial charges,
\eq{
\tilde Q = \frac{k}{2\pi\beta}\,\int\limits^\beta_0\extd\vp\,\eps^0{\mathcal X}^0\,,
}{eq:Qu1}
generate a centerless $\mathfrak{u}(1)$ current algebra off-shell. On-shell, \eqref{eq:34b} implies constancy of ${\mathcal X}^0$ so that only one generator remains. From \eqref{eq:29} we have
\begin{equation}
  \{Q\left(\varepsilon_1\right),Q\left(\varepsilon_2\right)\}= \frac{k}{2\pi\beta} \int\limits^\beta_0 \extd\varphi\,\delta_{\varepsilon_1}(\varepsilon_2 \mathcal{X}^0)=0\,,
  \label{eq:u1}
\end{equation}
once that $\delta_{\varepsilon_1}\varepsilon_2=0$ and $\delta\mathcal{X}^0=0$.

In order to get a non-trivial metric we no longer can use the group element \eqref{eq:b1}. Instead, we choose the group element in \eqref{eq:aca}, \eqref{eq:acx} as (inspired by the same choice in three dimensions \cite{Afshar:2016kjj})
\eq{
b=e^{\tfrac\rho2\,(L_+ - L_-)}\,.
}{eq:bu1}
The ensuing line-element is given by
\eq{
\extd s^2 = \extd\rho^2 + {\mathcal L}_0^2 \cosh^2\!\rho\,\extd\vp^2
}{eq:dsu1}
and the dilaton field reads
\eq{
X =  -\mathcal{X}^{0}\sinh\rho+(\mathcal{X}^{+}+\mathcal{X}^{-})\cosh\rho\,.
}{eq:Xu1}
The choices above, however, lead to a line-element \eqref{eq:dsu1} that is more naturally defined on the global AdS$_2$ strip rather than on the Poincar{\'e} disk. Thus, we do not discuss this case any further in the present work.

\section{Thermodynamics and entropy}\label{se:4}

In this section we consider aspects of thermodynamics, with particular focus on the entropy.
In section \ref{se:7.1} we derive entropy macroscopically, first by Wald's method, then from our on-shell action and finally from an asymptotically AdS$_2$ perspective. In section \ref{se:7.2} we show that naive applications of Cardy-like formulas gives results for entropy that agree with the macroscopic results of section \ref{se:7.1}.

\subsection{Wald's Method}\label{se:7.1}

To analyze static black holes let us perform a Wick rotation in the Euclidean line element \eqref{AdSLineElement} to Lorentzian signature.
With the definition $N\equiv h^2+(\mathcal L^0)^2$, the line element is given by
\begin{equation}
\extd s^2=\extd\rho^2+2\mathcal{L}^0\extd\rho \extd\varphi + N\extd\varphi^2.
\end{equation}
After the replacements $\varphi \to i\,\bar{\varphi}$ and $\anot \to -i \bar{\mathcal{L}}^{0}$, the $\rho$-coordinate of the horizon, i.e., the point where  the Killing vector $\partial_{\bar{\varphi}}$ becomes null, is determined by $N(\rho_h)=0$. This gives
\begin{equation}
e^{\rho_h} = \frac{1}{2\mathcal{L}^{+}}\left[\pm \bar{\mathcal{L}}^{0}\pm \sqrt{4\mathcal{L}^{+}\mathcal{L}^{-}+\left(\bar{\mathcal{L}}^{0}\right)^{2}}\right]\, ,\label{en01}
\end{equation}
which leads to the Hawking temperature
\begin{equation}
T = \frac{1}{\pi}\sqrt{-\frac{1}{4}\left(\anot\right)^{2}+\mathcal{L}^{+}\mathcal{L}^{-}}=\frac{\sqrt{ M}}{\pi}\,.
\label{eq:TM}
\end{equation}
The latter equality follows from the definition \eqref{eq:12} in the static case. Using the relation \eqref{eq:17} between the Casimir function and $M$, we obtain the relation
\begin{equation}\label{eq:StaticTemperature}
  T=\frac{\sqrt{\mathcal{C}}}{\pi\bar{Y}} \, ,
\end{equation}
which coincides with the regularity condition \eqref{eq:38}.

One way to compute the entropy is to use Wald's method \cite{Wald:1999wa}. In two-dimensional dilaton gravities this leads to the general result \cite{Gegenberg:1994pv}
\begin{equation}
S_{\textrm{\tiny Wald}} = k X_h\,,
\label{eq:sbh}
\end{equation}
where $X_h = \mathcal X^+ e^{\rho_h} + \mathcal{X}^- e^{-\rho_h}$ is the value of the dilaton at the horizon. Using \eqref{en01}, we obtain $X_h=2\sqrt{\mathcal C}$. This leads to the relation
\begin{equation}
S_{\textrm{\tiny Wald}} = 2 k\sqrt{\mathcal C} =  2 k \bar Y \sqrt{M} = 2k\pi \bar{Y} T
\label{eq:swald}
\end{equation}
between the entropy and the temperature compatible with the 3$^{\rm rd}$ law.

As a non-trivial check for the above, one may determine the entropy through the Euclidean path integral. In the saddle-point approximation the path integral is dominated by any smooth classical geometries that obey the boundary conditions. In our case, this will generically be global AdS space, since this is a smooth geometry for any temperature, and a Euclidean black hole with the appropriate mass. The relation between temperature and Casimir in the latter case was obtained in \eqref{eq:38}.
The on-shell action for the Euclidean black hole is
\begin{equation}
  \label{eq:39}
  \Gamma|_\ts{EOM}=-\frac{k \beta}{ \pi \bar{Y} }\,\Big(\mathcal{C} + \frac{1}{4}\,c_0 \Big) ,
\end{equation}
where we have included the contribution from the constant $c_0$ appearing in the second order action.\,\footnote{This constant produces a temperature-independent shift in the free energy. Therefore the first order action, which does not include this contribution, yields the same entropy.} The free energy is obtained by multiplying the on-shell action with the temperature
\eq{
F = -\frac{k}{\pi\bar Y}\,{\cal C} + F_0\qquad\qquad F_0 = \frac{k}{4\pi\bar{Y}}\,c_0\,.
}{eq:F}
The result that free energy scales linearly with the Casimir is quite universal for two-di\-men\-sio\-nal dilaton gravity and not a specific property of the JT model. The particular relation \eqref{eq:StaticTemperature} between the Casimir and temperature however is specific to JT and yields free energy as function of temperature.
\begin{equation}
F = -k\pi\bar{Y}T^2 + F_0
\end{equation}
Hence the entropy, defined as $S=-\partial F/\partial T$, is given by
\begin{equation}
S = 2k\pi \bar{Y}T\,.
\label{eq:S}
\end{equation}
This coincides with the computation of the entropy using Wald's method \eqref{eq:swald}.

A second check follows from the AdS$_2$ asymptotics, which allows us to construct the boundary stress tensor and compute the conserved charge associated with the static configuration Killing vector $\partial_{\varphi}$. Taking the boundary to be a surface of constant $\rho = \rho_c$, the leading term in the boundary metric as $\rho_c \to \infty$ is $\gamma_{\varphi\varphi} = e^{2\rho_c}(\ap)^{2}$. Then the single component of the boundary stress tensor is
\begin{gather}
   T^{\varphi\varphi} = \frac{2}{\sqrt{\gamma}}\,\frac{\delta \Gamma}{\delta \gamma_{\varphi\varphi}}\Big|_\ts{EOM} = e^{-3 \rho_c} \frac{k}{\pi}\,\Big(\mathcal{C} + \frac{1}{4}\,c_0\Big)\,\frac{1}{(\ap)^2 \xp} \, .
\end{gather}
Lowering the indices on $T^{\varphi\varphi}$ and contracting its indices with the Killing vector $\xi^{\varphi}=1$ and the unit-normal $u^{\varphi} = e^{-\rho_c}(\ap)^{-1}$ gives the internal energy
\begin{gather}
    E = \frac{k}{\pi}\,\Big(\mathcal{C} + \frac{1}{4}\,c_0\Big)\,\frac{\ap}{\xp} \, .
\end{gather}
Replacing $\ap/\xp = 1/\bar{Y}$ and using \eqref{eq:StaticTemperature}, the first law $\extd E = T\,\extd S$ gives 
\begin{gather}
    S = 2 k \sqrt{\mathcal{C}}= 2\pi k \bar{Y} T ~,
\end{gather}
in agreement with the other methods of calculating the entropy.

\subsection{Cardyology}\label{se:7.2}

Even though we do not have a Virasoro algebra on-shell as asymptotic symmetry algebra, we saw in sections \ref{se:2.3} and \ref{se:2.2} that a Virasoro algebra emerges when we are slightly off-shell, i.e., if we drop the on-shell condition of constancy of the Casimir. We can then cautiously use the Cardy formula to check whether it yields the correct black hole entropy derived macroscopically in section \ref{se:7.1}. The main input in the Cardy formula \eqref{eq:cardy} is the eigenvalue of the zero mode of the Virasoro algebra, which does have a canonical realization as generator of our asymptotic symmetry algebra even on-shell. So there is a chance that the Cardy formula works since we found a way to determine the value of the central charge $c$ in \eqref{eq:43b}.

The Cardy formula for a single Virasoro algebra is given by
\begin{equation}
\label{eq:cardy}
S_{\textrm{\tiny Cardy}}=2\pi\sqrt{\frac{c\mathcal{\bar{L}}_0}{6}}\,,
\end{equation}
where $c$ is the central charge and $\mathcal{\bar{L}}_0$ is the zero-mode of $\mathcal{L}$, rescaled suitably,
$\mathcal{\bar{L}}:=\frac{c}{12}\,\mathcal{L}$, $\mathcal{\bar{L}}_0:=\frac{1}{\beta}\int_0^\beta\extd\vp\,\mathcal{\bar{L}}$.
These definitions give a canonically rescaled version of the infinitesimal Schwarzian derivative \eqref{cbc01}
\begin{equation}
\delta_{\sigma}\mathcal{ \bar L} =\sigma{\bar{\mathcal{L}}}^\prime+2{\sigma}^\prime\mathcal{ \bar L}+\frac{c}{12}{\sigma}'''\,.
\end{equation}
Once we have the correct scaling for the zero-mode, we need the value of the central charge to determine the Cardy entropy \eqref{eq:cardy}. In \eqref{eq:43b} we derived an off-shell result for the central charge, $c=6k\bar Y/\pi$. Plugging this result into the Cardy formula \eqref{eq:cardy} and using the relations between $\mathcal{\bar{L}}$, $\mathcal{L}$, $M$ and $T$ [see Eqs.~\eqref{eq:ML}, \eqref{eq:TM}] yields
\eq{
S_{\textrm{\tiny Cardy}} = 2k\bar Y\sqrt{M} = 2\pi k\bar Y T\,,
}{eq:cardyology}
in agreement with the macroscopic result \eqref{eq:swald}. The result \eqref{eq:cardyology} shows that the Cardy formula works, which suggests that the off-shell central charge \eqref{eq:43b} is a meaningful quantity.\footnote{The result \eqref{eq:43b} for the central charge and the related Cardyology \eqref{eq:cardyology} was already presented in \cite{Grumiller:2015vaa, Grumiller:2013swa}. However, in the second paper, among other issues, the charges were non-integrable and in the first paper integrability was only achieved perturbatively. Moreover, rather than keeping fixed the zero mode of $1/Y$, the zero mode of the leading order function in the dilaton was fixed, which is equivalent only for zero mode solutions. Thus, the Cardyology in those papers was on shakier grounds than the one in the present work.}

The warped conformal case discussed in section \ref{se:2.2} allows to relate its entropy to the warped conformal analog of the Cardy formula. Using on-shell conditions the Wald entropy \eqref{eq:sbh} can be expressed entirely in terms of the Casimir $\cal C$ and the function $P_0$ defined in \eqref{eq:warped0}. If we set $P_0=0$ the analysis of section \ref{se:7.1} applies and we recover $S_{\textrm{\tiny wcbc}} = 2k\sqrt{\cal C}$, in agreement with the first equality in \eqref{eq:swald} and also in agreement with the Cardyology above. If $P_0$ is non-zero, but has only a zero-mode, i.e., $\partial_\vp P_0=0$, then we obtain
\eq{
S_{\textrm{\tiny wcbc}} = k\,X_h = 2k\,\sqrt{{\cal C}+P_0^2} = 2\pi\sqrt{\frac{k\bar Y}{\pi}\Big(\frac{k}{\pi\bar Y}({\cal C}+2P_0^2)-\frac{P^2_0 k}{\pi\bar Y}\Big)}\,.
}{eq:warped1}
The result \eqref{eq:warped1} looks like the warped conformal entropy \cite{Detournay:2012pc} (assuming $P_0^{\textrm{vac}}=0$)
\eq{
S_{\textrm{\tiny wCFT}} = 2\pi\sqrt{\frac{c}{6}\,\big(\hat L_0-\hat P_0^2/k_{\textrm{\tiny KM}} \big)}
}{eq:warped2}
where $\hat L_0$ and $\hat P_0$ are eigenvalues of the zero mode Virasoro and $\mathfrak{u}(1)$ generators, respectively, $c$ is the Virasoro central charge and $k_{\textrm{\tiny KM}}$ determines the $\mathfrak{u}(1)$ level. Comparing the two expressions \eqref{eq:warped1}, \eqref{eq:warped2} then leads to the matching conditions
\eq{
c=\frac{6k\bar Y}{\pi}\qquad\quad\hat L_0=\frac{k}{\pi\bar Y}({\cal C}+2P_0^2)=\frac{k\bar Y}{\pi}\,(M+2\hat{P}^2)\qquad\quad\frac{\hat P_0^2}{k_{\textrm{\tiny KM}}} = \frac{\big(\frac{k\bar Y}{\pi}\,\hat{P}\big)^2}{\frac{k\bar Y}{\pi}}
}{eq:warped3}
consistently with the result \eqref{eq:wcft3}. Note that for positive $k\bar Y$ both the central charge $c$ and the $\mathfrak{u}(1)$ level $k_{\textrm{\tiny KM}}$ are positive, compatible with unitarity.

\section{Conclusions}\label{se:5}

For a summary of our main results we refer to the introductory section \ref{se:1}. We conclude now with an outlook to possible further developments. 

The similarity of warped conformal boundary conditions in section \ref{se:2.2} to conformal ones in section \ref{se:2.3}, together with the relation of the latter to SYK, suggests the possibility of SYK-like models that exhibit an off-shell warped conformal symmetry that is largely broken on-shell. This may provide a new and interesting angle on SYK-like model building on the field theory side and lead to a generalized Schwarzian action along the lines of \cite{Barnich:2017jgw}.

Our focus in this paper was on the JT model, but the discussion in appendix \ref{app:B} makes it plausible that our analysis can be extended to fairly generic models of dilaton gravity in two dimensions. On general grounds, we expect a result for the free energy obtained from the Euclidean on-shell action analogous to \eqref{eq:F}, i.e.,
\eq{
F-F_0 \propto {\cal C} 
}{eq:concl1}
where $F_0$ is some state-independent constant and ${\cal C}$ is the Casimir function. While the result \eqref{eq:concl1} is essentially model-independent, the relation between Casimir and temperature will depend on the model. Besides doing such a general analysis (generalizing the one in \cite{Grumiller:2007ju} to situations where the dilaton fluctuates to leading order near the boundary) it will be of interest to discuss in detail specific selected models, such as spherically reduced Einstein gravity or other asymptotically flat dilaton gravity models. This may allow to find novel types of asymptotically flat boundary conditions in four or higher dimensions.

In generic models of Maxwell-dilaton gravity the zoology of holography branches out into more species than discussed in the present work, depending on the asymptotic behavior of the fields, which in turn are determined by the coupling functions depending on the dilaton. It could be of interest to generalize the analysis of \cite{Bagchi:2014ava}, which classified the models into asymptotically dilaton dominated, asymptotically confinement dominated and asymptotically mass-dominated, to looser fall-off conditions, similar to the ones considered in the present work but for more generic models.

\acknowledgments

DG is grateful to Max Riegler for collaboration on AdS$_3$ boundary conditions that inspired some of the developments in the present work.
We also thank Hern\'an Gonz\'alez for discussions.

This work was supported by the Austrian Science Fund (FWF), projects P~27182-N27 and P~28751-N27. DG and DV were additionally supported by the program Science without Borders, project CNPq-401180/2014-0. CV was supported by CNPq project 150407/2016-5 and CAPES post-doctoral scholarship. DV was supported by the grant 2016/03319-6 of S\~ao Paulo Research Foundation (FAPESP),  by the grants 303807/2016-4 and 456698/2014-0 of CNPq, and by the Tomsk State University Competitiveness Improvement Program. RM was supported by a Loyola University Chicago Summer Research Grant.

\appendix
\section{Modified bracket for diffeomorphisms}
\label{app:ModifiedBracket}

The functions appearing in the diffeomorphisms that generate asymptotic symmetries of the JT model are related to the parameters of the PSM large gauge transformation \eqref{eq:BoundaryGaugeTransformation} by
\begin{gather}\label{Identification:app}
	\lambda = - \vepsnot \qquad \sigma = - \frac{\vepsp}{\ap} \qquad \alpha = \frac{1}{\ap}\,\vepsm - \frac{\am}{(\ap)^{2}}\,\vepsp ~.
\end{gather}
In the PSM, the components of $\veps$ transform as $\delta_{\veps_1} \veps_{2} = [\veps_1,\veps_2]$, which corresponds to
\begin{align}\label{eq:App2}
	\delta_{\veps_1} (\sigma_2 \ap) = & \,\, \lambda_1\,\sigma_2\,\ap - \lambda_2\,\sigma_1\,\ap \\ \label{eq:App3}
	\delta_{\veps_1} (\lambda_2) = & \,\, 2\,(\ap)^{2}\,\big(\sigma_1\,\alpha_2 - \sigma_2\,\alpha_1\big) \\ \label{eq:App4}
	\delta_{\veps_1} (\sigma_2 \am - \alpha_2 \ap) = & \,\, \am\,(\lambda_2 \, \sigma_1 - \lambda_1 \, \sigma_2) - \ap (\lambda_2 \, \alpha_1 - \lambda_1 \, \alpha_2) ~.
\end{align}
But under the action of a diffeomorphism $\xi_1$, the components of $\xi_2$ transform according to the Lie derivative
	$\delta_{\xi_1} \xi_{2}^{\,\mu} = \pounds_{\xi_1} \xi_{2}^{\,\mu}= [\xi_1, \xi_2]^{\mu}$.
Writing the generators  $\xi$ of asymptotic symmetries in the form $\xi^{\mu} = \xi^{\ms{(0)}\mu} + e^{-2\rho}\,\xi^{\ms(2)\mu} + \ldots$, this bracket is
\begin{gather} \label{BracketTransformation1}
	[\xi_1,\xi_2]^{\mu} = \delta_{\xi_1} \xi_{2}^{\ms{(0)}\mu} + e^{-2\rho} \delta_{\xi_1} \xi_{2}^{\ms{(2)}\mu} + \ldots
\end{gather}
This gives the following for the action of $\xi_1$ on the functions appearing in $\xi_2$
\begin{align}
	\delta_{\xi_1} \sigma_2 = & \,\, \sigma_{1} \, \sigma_{2}' - \sigma_{2} \, \sigma_{1}' \\
	\delta_{\xi_1} \lambda_2 = & \,\, \sigma_{1} \, \lambda_{2}' - \sigma_{2} \, \lambda_{1}' + \sigma_{2} \, \big(\lambda_{1}' - 2\,(\ap)^{2} \alpha_{1}\big) \\
	\delta_{\xi_1} \alpha_2 = & \,\, \sigma_{1}\,\alpha_{2}' - \sigma_{2} \, \alpha_{1}' + \alpha_{1} \, \sigma_{2}' - \alpha_{2} \, \sigma_{1}' + 2 \alpha_{1} \big(\lambda_{2} - \anot\,
	\sigma_{2}\big)  - 2 \alpha_{2} \big(\lambda_{1} - \anot\,
	\sigma_{1}\big) ~.
\end{align}
These do not agree with the transformations from the PSM, even after accounting for the factors of $\apm$ in \eqref{eq:App2}-\eqref{eq:App4}. The reason for this apparent discrepancy is the explicit dependence of the functions $\sigma$, $\lambda$, and $\alpha$ on the functions appearing in the metric. So we need a modified bracket that isolates the parts of the functions appearing in $\xi^{\mu}$ that do not depend on $\apm$ and $\anot$, see for instance \cite{Barnich:2010eb, Compere:2015knw, Grumiller:2016pqb}.

To see how this works, consider a specific component of $\xi^{\mu}$ and write it as
\begin{gather}
	\xi^{\mu} = \veps^{A}\,F_{A}^\mu(g) + e^{-2\rho}\,\veps^{A}\,H_{A}^\mu(g) ~,
\end{gather}
where the index $A$ runs over $\pm,0$, and the functions $F_{A}^\mu$ and $H_{A}^\mu$ encode the dependence of the functions $\sigma$, $\lambda$, and $\alpha$ on the metric functions $\apm$ and $\anot$. Then
\begin{align}
	[\xi_1,\xi_2]^{\mu} = &\,\, \xi_{1}^{\nu} \, \partial_{\nu} (\veps_{2}^{A}\,F_{A}^\mu(g) + e^{-2\rho}\,\veps_{2}^{A}\,H_{A}^\mu(g)) - \xi_{2}^{\nu} \, \partial_{\nu} (\veps_{1}^{A}\,F_{A}^\mu(g) + e^{-2\rho}\,\veps_{1}^{A}\,H_{A}^\mu(g)) \nonumber \\
	= & \,\, F_{A}^\mu\,\big( \xi_{1}^{\nu} \, \partial_{\nu} \veps_{2}^{A} - \xi_{2}^{\nu} \, \partial_{\nu} \veps_{1}^{A} \big) + H_{A}^\mu\,\big( \xi_{1}^{\nu} \, \partial_{\nu} (e^{-2\rho} \veps_{2}^{A}) - \xi_{2}^{\nu} \, \partial_{\nu} (e^{-2\rho} \veps_{1}^{A}) \big) \nonumber \\ 
	& \,\, \quad + \veps_{2}^{A}\,\xi_{1}^{\nu} \partial_{\nu} F_{A}^\mu + e^{-2\rho}\,\veps_{2}^{A} \xi_{1}^{\nu} \partial_{\nu} H_{A}^\mu - \veps_{1}^{A}\,\xi_{2}^{\nu} \partial_{\nu} F_{A}^\mu - e^{-2\rho}\,\veps_{1}^{A} \xi_{2}^{\nu} \partial_{\nu} H_{A}^\mu ~.
\end{align}
The terms in the last line capture the part of the action of $\xi_1$ on $\xi_2$ that is due to the explicit metric-dependence of both vectors appearing in the bracket. We denote them by
\begin{gather}
	\delta_{\xi_1}^{g} \xi_{2}^{\mu} = \veps_{2}^{A}\,\xi_{1}^{\nu} \partial_{\nu} F_{A}^\mu + e^{-2\rho}\,\veps_{2}^{A} \xi_{1}^{\nu} \partial_{\nu} H_{A}^\mu ~.
\end{gather}
Then for this component of $\xi^{\mu}$ we have
\begin{align}
	[\xi_1,\xi_2]^{\mu} = & \,\, F_{A}^\mu\,\big( \xi_{1}^{\nu} \, \partial_{\nu} \veps_{2}^{A} - \xi_{2}^{\nu} \, \partial_{\nu} \veps_{1}^{A} \big) + H_{A}^\mu\,\big( \xi_{1}^{\nu} \, \partial_{\nu} (e^{-2\rho} \veps_{2}^{A}) - \xi_{2}^{\nu} \, \partial_{\nu} (e^{-2\rho} \veps_{1}^{A}) \big)  + \delta_{\xi_1}^{g} \xi_{2}^{\mu} - \delta_{\xi_2}^{g} \xi_{1}^{\mu} ~.
\end{align}
Following \cite{Barnich:2010eb}, we define the modified bracket $[\cdot,\cdot]_\ts{M}$ as the regular bracket with the contributions due to the metric dependence of $\xi_1$ and $\xi_2$ removed
\begin{gather}
	[\xi_1, \xi_2]_\ts{M}^\mu = [\xi_1, \xi_2]^\mu - \delta_{\xi_1}^{g} \xi_{2}^{\mu} + \delta_{\xi_2}^{g} \xi_{1}^{\mu} ~.
	\label{eq:bracket}
\end{gather}
Then, as in \eqref{BracketTransformation1}, the modified bracket defines the transformations for the parts of $\xi^{\mu}$ that are independent of $\apm$ and $\anot$ as
\begin{gather} \label{ModifiedBracketTransformation}
	[\xi_1, \xi_2]_\ts{M}^\mu = F_{A}^\mu(g) \, \delta_{\veps_1} \veps_{2}^{A} + e^{-2\rho}\,H_{A}^\mu(g) \, \delta_{\veps_1} \veps_{2}^{A}  ~,
\end{gather}
which should agree with the expected results from the PSM. We check this below.

Using the identification \eqref{Identification:app} and the transformations \eqref{DiffeoTransformhp} we have
\begin{align}
	- \delta_{\xi_1}^{g} \sigma_2 + \delta_{\xi_2}^{g} \sigma_1 = & \,\, \sigma_2\,\lambda_1 - \sigma_1\,\lambda_2 + \sigma_2\,\sigma_{1}' - \sigma_{1}\,\sigma_{2}' \\
	- \delta_{\xi_1}^{g} \lambda_2 + \delta_{\xi_2}^{g} \lambda_1 = & \,\, 0 \\
	- \delta_{\xi_1}^{g} \alpha_{2} + \delta_{\xi_2}^{g} \alpha_{1} = & \,\, \alpha_2\,\lambda_1 - \alpha_1 \, \lambda_2 + 2\,\anot\,(\alpha_1\,\sigma_2 - \alpha_2\,\sigma_1) + \alpha_2\,\sigma_{1}' - \alpha_{1}\,\sigma_{2}'  \nonumber \\
	&\,\, \quad + \sigma_{2}\,\alpha_{1}' - \sigma_{1}\,\alpha_{2}' - 2\,\frac{\am}{\ap}\,(\sigma_1\,\lambda_2 - \sigma_2\,\lambda_{1}) ~.
\end{align}
Now, using these results and the regular bracket for the components of $\xi$, we obtain
\begin{align} \label{ModifiedBracketPhi}
	[\xi_1,\xi_2]_\ts{M}^{\varphi} = & \,\, \sigma_{2}\,\lambda_{1} - \sigma_{1}\,\lambda_{2} + e^{-2\rho}\left[\alpha_{1}\,\lambda_{2} - \alpha_{2}\,\lambda_{1} - 2\,\frac{\am}{\ap}\,\big(\sigma_{1}\,\lambda_{2} - \sigma_{2}\,\lambda_{1} \big)\right] \\ \label{ModifiedBracketRho}
	[\xi_1,\xi_2]_\ts{M}^{\rho} = & \,\, 2\,(\ap)^{2}\,\big(\sigma_{1}\,\alpha_{2} - \sigma_{2}\,\alpha_{1} \big) - \anot\,[\xi_1,\xi_2]_\ts{M}^{\varphi} ~.
\end{align}
For the leading term in the component $\xi^{\varphi}$ we have $F_{+} = - 1/\ap$. Then \eqref{ModifiedBracketTransformation} and \eqref{ModifiedBracketPhi} give
\begin{gather}
	\delta_{\veps_{1}} ( \sigma_{2}\,\ap) =  \lambda_{1}\,\sigma_{2}\,\ap - \lambda_{2}\,\sigma_{1}\,\ap
\end{gather}
in agreement with \eqref{eq:App2}. For the $\mathcal{O}(e^{-2\rho})$ term in $\xi^{\varphi}$ the functions $H_{A}$ are $H_{-} = 1/\ap$ and $H_{+} = - \am / (\ap)^{2}$, and we have
\begin{gather}
	\frac{1}{\ap}\,\delta_{\veps_1} \veps_{2}^{-} - \frac{\am}{(\ap)^{2}}\,\delta_{\veps_1} \veps_{2}^{+} = \alpha_{1}\,\lambda_{2} - \alpha_{2}\,\lambda_{1} - 2\,\frac{\am}{\ap}\,\big(\sigma_{1}\,\lambda_{2} - \sigma_{2}\,\lambda_{1} \big) ~,
\end{gather}
which yields
\begin{gather}
	\delta_{\veps_{1}}\big(\sigma_{2} \, \am - \alpha_{2} \, \ap \big) = \am\,\big( \la_2\,\sigma_{1} - \la_1\,\sigma_{2} \big)  -\ap\big(\la_2\,\alpha_{1} - \la_1\,\alpha_{2}\big) 
\end{gather}
in agreement with \eqref{eq:App4}. Finally, for $\xi^{\rho}$, we have $F_{0} = -1$ and $F_{+} = \anot / \ap$. Then \eqref{ModifiedBracketTransformation} and \eqref{ModifiedBracketRho} give
\begin{gather}
	\delta_{\veps_{1}} (\lambda_{2}) = 2\,(\ap)^{2}\,\big(\sigma_{1}\,\alpha_{2} - \sigma_{2}\,\alpha_{1} \big) 
\end{gather}
in agreement with \eqref{eq:App3}.

In conclusion, the transformations obtained from the modified bracket \eqref{eq:bracket} give exactly the results expected from the PSM. Thus, equipped with the modified bracket \eqref{eq:bracket}, all the results in the second order formalism agree with those of the PSM.

\section{Toy models (batteries included)}\label{app:B}

In this appendix we address some aspects of boundary conditions, variational principle, averaged boundary charges, asymptotic symmetries etc.~for simple toy models that nevertheless share all key features of generic PSMs. This may help to elucidate some potentially confusing points encountered in the main text and allows to disentangle features that are specific to the JT model from more generic features of PSMs.

The first toy model studied in section \ref{app:B1} is abelian $BF$-theory in two dimensions and the second one the simplest symplectic special case of a PSM. In both models we address particularly the issue of varying ``chemical potentials''. In the concluding section \ref{app:B3} we address implications for generic dilaton gravity.

\subsection{Abelian \texorpdfstring{$BF$}{BF}-theory as Casimir sector of generic Poisson sigma models}\label{app:B1}

\newcommand{\oi}{\oint_{\partial\mathcal{M}}}

Consider the bulk action of abelian $BF$ theory on a disk,
\eq{
\Gamma_{BF}=-\frac{k}{2\pi}\int_{\mathcal{M}} B F
}{eq:app1}
with $F=\extd A$. The equations of motion are
\eq{
B=B_0=\textrm{const.}\qquad F=0\,.
}{eq:app2}
The variational principle is well-defined provided the zero mode of $A$ is not allowed to vary,
\eq{
\delta\Gamma_{BF}\big|_{\textrm{\tiny EOM}} = -\frac{k}{2\pi}\,B_0\oi\delta A = 0
}{eq:app3}
where the boundary integral is along the $S^1$ of the disk. Since the non-zero modes of $\delta A$ integrate to zero and the zero mode of $A$ is not allowed to vary the last equality holds, which establishes a well-defined variational principle.

Like in the main text we interpret the coordinate along the $S^1$, denoted by $\vp$, as Euclidean time; the radial coordinate of the disk is denoted by $ \rho$. From a canonical perspective the quantities $B$, $A_\rho$ and $A_\vp$ are canonical coordinates, momenta and Lagrange multipliers, respectively. The latter are often interpreted as chemical potentials. However, the condition $\delta A_\vp\neq 0$ (only for non-zero modes) implies that the standard interpretation of $A_\vp$ as ``chemical potential'' has to be taken with a few grains of salt --- after all, by ``chemical potential'' one usually means quantities $\mu$ that are arbitrary but fixed, and not quantities that are allowed to vary. However, the reason for doing so is usually that otherwise the variational principle would be ill-defined; for instance, in holographic contexts the first variation of the action schematically reads on-shell $\delta\Gamma\sim (\textrm{vev})\, \delta(\textrm{source})\sim Q\, \delta\mu$ so that a well-defined variational principle requires that the sources (or chemical potentials) are fixed, $\delta\mu=0$. By contrast, in abelian $BF$-theory no such condition is necessary, so we refrain from demanding $\delta A_\vp=0$ except for its zero mode.

The canonical boundary currents are given by
\eq{
\delta Q[\la]=\frac{k}{2\pi}\, \la\, \de B\,.
}{eq:app4}
Their integrated (both in field space and along the $S^1$) version reads
\eq{
\hat Q = \oi Q[\la] = \frac{k}{2\pi}\, B_0 \oi \la\,.
}{eq:app5}
This result implies that only the zero mode of the transformation parameter $\la$ generates an asymptotic symmetry, while all other Fourier modes of $\la$ generate pure gauge symmetries. 

The analysis above implies that in a PSM each Casimir function generates a single canonical boundary charge, namely the Casimir itself, which on-shell is constant.

\subsection{Darboux sector of generic Poisson sigma model}\label{app:B2}

Through target space diffeomorphisms it is always possible to bring a PSM into Casimir--Darboux coordinates, see for instance \cite{Strobl:1999wv}. We have dealt with the Casimir sector in section \ref{app:B1} above. Now we consider the Darboux sector (or symplectic sector).

To this end consider the simplest invertible Poisson tensor,
\eq{
P^{IJ}=\epsilon^{IJ}\qquad I,J\in\{1,2\}\qquad \epsilon^{12}=+1\,.
}{eq:app6}
The action is given by
\eq{
\Gamma_D = -\frac{k}{2\pi}\,\int_{\mathcal M}\big[X^I\extd A_I + \tfrac12\epsilon^{IJ}A_I\wedge A_J\big] + \Gamma_{\partial\mathcal M}
}{eq:app7}
with the boundary action
\eq{
\Gamma_{\partial\mathcal M}= \frac{k}{2\pi}\,\oi X^1A_1 \,.
}{eq:app8}
Choices different from \eqref{eq:app8} are possible, but we focus here on this particular example. Regarding the coordinates and interpretation of the fields the same remarks apply as in section \ref{app:B1} above.

The equations of motion 
\eq{
F_I=\extd A_I=0\qquad \extd X^I + \epsilon^{IJ} A_J = 0
}{eq:app9}
can be solved for $A_I$ in terms of $X^I$,
\eq{
A_{1,2} = \pm \extd X^{2,1}\,.
}{eq:app10}
The action of gauge transformations on the fields
\eq{
\delta_\la X^{1,2}=\pm\la_{2,1}\qquad \de_\la A_{1,2}=-\extd\la_{1,2}
}{eq:app11}
is of course compatible with the solutions \eqref{eq:app10}.

The first variation of the action reads on-shell
\eq{
\delta\Gamma_D\big|_{\textrm{\tiny EOM}} = \frac{k}{2\pi}\,\oi\big[A_1\, \delta X^1 - X^2\, \delta A_2\big]\,.
}{eq:app12}
Now we need to restrict the variations or fields in the integrand of \eqref{eq:app12} such that the integral vanishes identically. There are various ways of doing this; for instance, one could impose the rather strong conditions $\delta A_2=A_1=0$. Instead, we impose only one condition on the variation of the ``chemical potentials'' $A_{I\vp}$, namely that one of these variations satisfies the on-shell condition
\eq{
\delta A_{2\vp} = - \partial_\vp\, \delta X^1\,.
}{eq:app13}
This choice implies that $X^{1,2}$ and $A_{1\vp}$ are allowed to fluctuate arbitrarily. Again, we have the situation that the ``chemical potentials'' $A_{I\vp}$ are not fixed; rather, one of them has to fluctuate to maintain the on-shell condition \eqref{eq:app13} while the other one can fluctuate freely. Moreover, as a consequence of the choices \eqref{eq:app8} and \eqref{eq:app13} the variational principle is well-defined,
\eq{
\delta\Gamma_D\big|_{\textrm{\tiny EOM}}= \frac{k}{2\pi}\,\oi \extd\big(X^2\,\delta X^1\big) = 0
}{eq:app14}
since the term in the integrand of \eqref{eq:app14} is a total derivative.

The integrated (again, in field space and along the $S^1$) boundary charges are given by
\eq{
\hat Q[\la_I] = \frac{k}{2\pi}\,\oi \big[\la_1 X^1 + \la_2 X^2\big]
}{eq:app15}
so that now we have two infinite towers of such charges, namely all Fourier modes of $X^1$ and $X^2$. The asymptotic symmetry algebra is given by
\eq{
\{\hat Q[\la_I],\,\hat Q[\la_J]\} = \de_{\la_J} \hat Q[\la_I] = \epsilon^{IJ}\, \frac{k}{2\pi}\, \oi \la_I\la_J \,.
}{eq:app16}
The algebra \eqref{eq:app16} is the loop algebra of the Heisenberg algebra (or, equivalently, infinite copies of the Heisenberg algebra).

It is instructive to check what happens if we drop the boundary term \eqref{eq:app8}. Then the variational principle is well-defined if $\delta A_{I\vp}=0$, since $\delta\Gamma|_{\textrm{\tiny EOM}}\propto \oi X^I\delta A_I$. This means that $A_{I\vp}$ are geniuine chemical potentials. Boundary condition preserving transformations are then restricted by the condition
\eq{
\de_\la A_{I\vp} = -\partial_\vp \la_I = 0
}{eq:app17}
which implies that the transformation parameters $\la_I$ have vanishing non-zero modes. As a consequence, there are only two boundary charges rather than two infinite towers of them, namely the zero modes of $X^I$. 

The toy model above clearly shows how crucial it can be to relax the condition that Lagrange multipliers like $A_{I\vp}$ are fixed: if they are fixed the asymptotic symmetry algebra consists of a single copy of the Heisenberg algebra, while the relaxed condition \eqref{eq:app13} [together with the required boundary term \eqref{eq:app8}] lead to an enhancement to the loop group of the Heisenberg algebra. This is precisely what we found also for the loop group boundary conditions in section \ref{se:2.1}.

\subsection{Implications for dilaton gravity}\label{app:B3}

The toy models in the previous sections are physically not too rich. However, adding both actions \eqref{eq:app1} and \eqref{eq:app7} leads to an interesting PSM with 3-dimensional target space in Casimir--Darboux coordinates. The reason such PSMs are of physical interest is their relation to generic models of two-dimensional dilaton gravity. In terms of the target space coordinates used in section \ref{se:2}, $X$, $\hat X^\pm = X^{\hat 0} \pm X^{\hat 1}$, the Casimir--Darboux target space coordinates $X^I_{\textrm{\tiny CD}}$ are given by (see e.g.~\cite{Strobl:1999wv})
\eq{
X^I_{\textrm{\tiny CD}} = \{\mathcal{C},\,X,\,\ln|X^+|\}\,.
}{eq:app18}
Here we have assumed Lorentzian signature and imposed the condition $X^+\neq 0$ (geometrically this restricts to an Eddington--Finkelstein patch, which is more than sufficient for our purposes). For sake of specificity we further restrict to $X^+>0$. The physical interpretation of the target space coordinates is mass function, dilaton field and Lorentz angle, respectively. The target space coordinates defined in \eqref{eq:app18} obey 
\eq{
\{\mathcal{C},\, X^I_{\textrm{\tiny CD}}\} = 0 \qquad \{X,\,\ln|X^+|\} = 1
}{eq:app19}
which are nothing but the Poisson brackets for Casimir--Darboux coordinates, with the Casimir $\cal C$ and the Darboux coordinates $X$ and $\ln|X^+|$.

If the Casimir function $\mathcal C$ is considered as the Hamilton operator of the theory then all excitations generated by the canonical boundary charges are ``soft'' in the sense that they commute with this Hamiltonian. Thus, PSMs in Casimir--Darboux coordinates with the boundary conditions discussed in section \ref{app:B2} behave similar to three-dimensional gravity with soft Heisenberg hair \cite{Afshar:2016wfy, Afshar:2016kjj}. A key difference is that in the three-dimensional case there are two Casimirs instead of one.

In conclusion, besides illuminating certain technical aspects of the discussion in the main text, the analysis of this appendix suggests that even for generic two-dimensional dilaton gravity models it could be possible to obtain infinite dimensional asymptotic symmetry algebras, in particular infinite copies of the Heisenberg algebra supplemented by one Casimir.


\providecommand{\href}[2]{#2}\begingroup\raggedright\endgroup

\end{document}